# Photosynthetically-powered phototactic active nematic fluids and gels

**Authors:**

Andrii Repula[1], Colin Gates[2,3], Jeffrey C. Cameron[2,3] and Ivan I. Smalyukh[1,2,4,5]*

**Affiliations:**

[1]Department of Physics, University of Colorado, Boulder, Colorado 80309, USA

[2]Renewable and Sustainable Energy Institute, National Renewable Energy Laboratory and University of Colorado, Boulder, Colorado 80309, USA

[3]Department of Biochemistry, University of Colorado, Boulder, Colorado 80309, USA

[4]International Institute for Sustainability with Knotted Chiral Meta Matter (WPI-SKCM²), Hiroshima University, Higashihiroshima, Hiroshima, 739-8526, Japan

[5]Materials Science and Engineering Program and Department of Electrical, Computer and Energy Engineering, University of Colorado, Boulder, Colorado 80309, USA

*Corresponding author. Email: ivan.smalyukh@colorado.edu

**Abstract.** One of the most ancient forms of life dating to ~3.5 billion years ago, cyanobacteria are highly abundant organisms that convert light into energy and motion, often within conjoined filaments and larger colonies. We study how gradients of light intensity trigger orderly phototactic motions and dense bacterial communities, which remained quantitatively unexplored despite being among the oldest forms of active living matter on Earth. The phototaxis drives a transition from initially polar motions of semiflexible long filaments along complex curved spatiotemporal trajectories confined within illuminated areas to their bipolar motility in the ensuing crowded environment. We demonstrate how simply shining light causes a spontaneous self-assembly of two- and three-dimensional active nematic states of cyanobacterial filaments, with a plethora of motile and static topological defects. We quantify light-controlled evolutions of orientational and velocity order parameters during the transition between disordered and orientationally ordered states of our photosynthetic active matter, as well as the subsequent active nematic's fluid-gel transformation. Patterned illumination and foreign inclusions with different shapes interact with cyanobacterial active nematics in nontrivial ways, while inducing soft interfacial boundary conditions and quasi-boojum-like defects. Commanding this cyanobacterial collective behavior could aid inhibiting generation of toxins or enhancing production of oxygen and biomaterials.

Active matter, which includes all forms of life and their motile building blocks, as well as many synthetic systems, is characterized by its ability to take in and dissipate energy at the level of constituent particles and, in the process, execute systematic motions[1-3]. This out-of-equilibrium behavior often leads to emergent collective phenomena like the formation of flocks of birds and schools of fish[4,5], differing from more conventional equilibrium condensed matter systems[6-11]. Formation of and transitions between ordered states of active matter with polar or nonpolar motions of constituents are of particular interest, but systems in which such behaviors can be systematically studied and robustly controlled remain limited[12-22]. For instance, the control and re-supply of various common chemical sources of energy, including nutrients, can be limited and become a factor in observing the active matter behavior in classical active matter systems like bacterial communities, including the limited guiding of their behavior via phenomena like chemotaxis[9-11,23]. On the other hand, cyanobacteria use light as an energy source derived through photosynthetic activity, which once transformed our Earth's atmosphere while generating oxygen and converting carbon dioxide into biomass[24-30]. They are also the first bacteria to be discovered and directly observed in a microscope, with the very first historical microbiology reports mentioning the light-powered activity of filaments consisting of "green globules joined together" and moving "in an orderly manner"[31]. Such light-powered active matter systems may offer unprecedented control of the out-of-equilibrium behavior, potentially even in relation to mats, blooms and production of oxygen or toxins[24-30]. Surprisingly, active matter behavior of cyanobacterial filaments remained quantitatively unexplored, with the great potential of controlling it by light and gradients of its intensity never utilized. Exploring and exploiting this behavior is fundamentally important to understand, for example, how complex filamentous cyanobacterial motility[32-34] changes upon formation of a crowded orientationally ordered environment and how defects can form during light-driven "fusion" of nematic domains with different orientations, an active-matter analog of the Kibble-Zurek mechanism[35-37]. The possibility of forming light-induced active nematic counterparts of tactoidal domains with possible analogs of boojum defects, which are common for co-existent nematic and disordered states in passive liquid crystals[38-43], also remains unexplored despite of the tremendous recent progress in studies of defects in various biological and synthetic active matter systems[1-20,44,45]. Even the very emergence of active nematic order out of a disordered state in response to light has not been demonstrated, even though this would be a fascinating active matter analog of the phototropic liquid crystallinity[46]. Despite of the growing interest in using light to control active matter,[47-49] as well as despite of the related to it great potential for realizing anomalous stimuli-responsive materials,[50] the simultaneously photosynthetic (powered by light as the source of energy) and phototactic contiguous active nematic liquid crystals have not been studied.

Here we use two different species of cyanobacteria, *Geitlerinema* sp. and *Oscillatoria brevis* (*O. brevis*), to reveal how phototaxis and spatially localized light-supplied energy lead to spontaneous formation of active liquid crystal states, where both two-dimensional (2D) and three-dimensional (3D) active nematic fluids and gels form under different conditions. In this context, we study the transition from initially predominantly polar curvilinear motions of long semiflexible conjoined cyanobacterial filaments within illuminated areas to their emergent bipolar motility as the system transforms from a collection of sparce non-interacting filaments to the nematic-state crowded cyanobacterial community. Contrary to naïve expectations, such effects do not emerge upon concentrating the cyanobacterial filaments by simple physical means like centrifugation instead of phototaxis, where separation of bacterial filaments is not controlled by natural slime production, becoming detrimental for the bacterial community's viability and highlighting the importance of our methodology of obtaining phototactic active nematics compatible with natural microbiological processes. By quantitatively characterizing mean square displacements (MSDs) and both orientational and nematic velocity order parameters, we probe both disordered and orientationally ordered active matter fluid and gel states, as well as transformations between them. We find that both motile and static topological defects emerge in these active nematic systems, with their dynamics sensitive to the fluid or gel types of the photosynthetically powered nematic host media. Furthermore, we explore interactions of the locally induced active nematic's orientational order and bacterial motility with edges of illuminated sample areas and with immobile foreign inclusions, showing emergence of quasi-boojum-like defects. Our findings may allow for designing highly controlled experiments to test active matter theories as well as may provide insights for commanding the out-of-equilibrium collective behavior of cyanobacterial active matter,[50] with potential utility including control of bacterial mats and blooms, as well as oxygen generation and inhibition of toxin production[24-30].

**Results**

**2D phototactic active nematics**

Cyanobacteria are photosynthetic organisms that exhibit positive phototactic motility toward localized light sources (Fig. 1a-c), with the light being the source of energy. They tend to live colonially within long filaments of tens to hundreds of rod- or disc-shaped cells, called trichomes (Fig. 1b and Supplementary Fig.1). Resembling stacks of molecules in chromonic liquid crystals[11], trichomes naturally have large length-to-width aspect ratios, typically ~100 for the species we study (though we can control it, see Methods). They can exhibit either bent or straight shapes that often correlate with circular or straight motion trajectories, with some species also rotating around longitudinal axes while self-propelling (Supplementary Fig. 2 and Movies 1 and 2). Cyanobacteria secrete polysaccharide slime[32] (Supplementary Fig. 1d) to mediate gliding motion along the filament axis, which is further enriched by the type IV pili[33]. Once in the illuminated region (Fig. 1b), cyanobacterial motility becomes geometrically confined to it (Fig. 1c) due to the spatially nonuniform availability of locally abundant energy supplied as light within the illuminated area. While the phototaxis-driven motion from dark to lit areas is directed towards the optimally illuminated region and polar in nature for all studied cyanobacterial filaments, to retain the light-supplied access to energy by staying in the lit area, they are found exhibiting different, more complex motions either along complex curvilinear trajectories or with motion directionality reversals, depending on trichomes' effective rigidity, geometry and species (Supplementary Fig. 2). Although cell divisions are negligible under the relatively short-term conditions of our experiments[30], filament number density in the lit area increases with time owing to migration of cells from darker regions, which in turn leads to out-of-equilibrium inter-particle interactions via collisions (Supplementary Movie 1).

When agar-enabled confinement constrains cyanobacterial displacements to be within a one-bacterium-thick 2D "monolayer"[30] (see Methods), collision between trichomes results in their mutual rotations and co-alignment (Fig. 1d,e, and Supplementary Movie 1). The phototaxis-driven increase of cell density over time leads to the emergence of out-of-equilibrium nematic order of the active cyanobacterial particles within the entire illuminated area (Fig. 2a-e, Fig. 3a-c and Supplementary Movies 2 and 3). These observations vividly resemble the onset of interactions that in simple early models of active rods[8] lead to 2D nematic order, though (at the time of moving towards the lit areas) individual trichomes exhibit polar motility in the process of phototaxis, which they can "switch to bipolar". As we shall see later, overall this active matter system exhibits a very complex, hierarchically emergent behavior that cannot be fully captured by theories known to us so far. The ensuing 2D nematic is characterized by a director **n** describing the average orientation of cyanobacterial filaments (Fig. 2b), which is selected out of all possible azimuthal orientations as a result of inter-trichome interactions. The 2D active nematics emerge under different conditions of the 2D monolayer-like confinement, both atop solid surfaces like glass (Fig. 2b and e) and at water-air interfaces (Fig. 2d and Supplementary Movie 4), with the latter being a 2D laboratory counterpart of blooms and cyanobacterial mats found on surfaces of freshwater lakes[28]. In both examples phototaxis drives out-of-equilibrium structural transformations from an initial isotropic state with the active particles moving in random directions at their low concentrations or number densities (Fig. 2a,c) to a nematic state with orderly bipolar motility at high number densities (Fig. 2b,c), as quantified experimentally using video microscopy and data analysis (Supplementary Fig. 3 and Movies 2,3).

As the emergent nematic order develops, the cyanobacterial cell motility evolves dramatically over time. Initially predominantly polar motions of long filaments towards the light source or along curved trajectories confined within the illuminated region (Fig. 1c and Supplementary Fig. 2a) gradually change to their bipolar motility, which is first oriented along positionally random but orientationally ordered trajectories (nematic-fluid-like) and then (in older samples in a state to which we will refer as a "nematic gel") features bacteria repeatedly moving along the same slime-defined trajectories in a bipolar fashion, as we will quantify further below. Interestingly, while motility features and complex propulsion trajectories of planktonic cyanobacterial filaments highly depend on their length-to-width aspect ratio, rigidity of the semi-flexible trichomes and their type/species (Supplementary Fig. 2), all studied trichomes are found to eventually form phototaxis-driven active nematic fluid and gel states with bipolar motility in the crowded nematic environment. This can be explained, at a qualitative level, by the semi-flexible trichomes adopting straight conformations to maximize the surface coverage of the lit areas and, thus, to optimize the energy intake by the growing dense bacterial community. Even as we start from relatively short straight trichomes (Fig. 2a, prepared with the help of sonication, see Methods), we find them becoming longer and longer due to end-to-end associations as they form nematic states (Fig. 2b) but still retaining straight-rod-like conformations and featuring bipolar motility along straight lines decorating the nonpolar director field $\mathbf{n}(\mathbf{r})$ (Fig. 2c).

Video-microscopy reveals how, over the course of hours, orientational distributions of bacterial filaments evolve from isotropic to unidirectional nematic whereas velocity vector distributions evolve from initially pointing in random directions in the isotropic state to adopting a bimodal character in the nematic state (Fig. 2c and f, Supplementary Movie 2). For example, we quantitatively characterize evolution of the system within 8 hours (Fig. 2g-i). For this, we concurrently measure the relative filament surface coverage area $\sigma$ (the ratio of the surface area covered by filaments divided by the total area), the 2D orientational order parameter[6] $S_{2D}=<2cos^2\theta-1>$ and the 2D velocity order parameter[7] $\varphi = <\varphi(t)>_t = 2((<cos^2\chi>_t - 1/2)^2 + <sin\chi cos\chi>^2_t)^{1/2}$ (Fig. 2g-i), where angles $\theta$ and $\chi$ describe the individual trichome's and its velocity vector's orientations, respectively (Fig. 2f). While the behavior of cyanobacterial active particles is species-specific and much more complex than that of active rigid rods (Supplementary Fig. 2), their behavior in the crowded cyanobacterial community environment resembles that of rigid rods with bipolar motility and justifies the use of $\varphi$ and $S_{2D}$ as quantifiers of the observed emergent active matter behavior, even if the physics behind these effects is complex. Starting from a dilute cyanobacterial community with $\sigma \approx 0.02$ and filament average length of 40 μm (Supplementary Fig.1), local illumination prompts continuous growth of surface coverage until it reaches a saturation level of $\sigma \approx 0.5$ in about 3 hours (Fig. 2g). Both orientation and velocity order parameters continuously increase up to $S_{2D} \approx 0.70$ (Fig. 2h) and $\varphi \approx 0.84$ (Fig. 2i), also saturating in ~3 hours. Videomicroscopy with individual cyanobacterial trichome tracking reveals the bipolar propulsion mainly parallel to **n** (Supplementary Movie 2), with MSD found scaling as $t^\alpha$ (Fig. 2j), where $\alpha$ is the diffusion exponent[6]. At short time scales, the propulsion along **n** is essentially ballistic for both nematic fluids and gels, with $\alpha_\parallel \approx 2$, similar to the case of ballistic propulsion of individual cyanobacterial cells powered by light (Supplementary Fig. 1g). At the same time, the diffusion exponent measured perpendicular to **n** in the ordered crowded communities is hindered, featuring $\alpha_\perp < 2$ (Fig. 2j). These findings are consistent with propulsion forces being along the trichome's long axis. Interestingly, at short time scales (shorter than the typical time of reversing propulsion directionality) the diffusion exponents are relatively similar for active nematic fluids and gels, implying that the filaments continue to exhibit active particle behavior during both transformations from isotropic to nematic fluid state and nematic fluid to gel state (Fig. 2j), though below we shall see the differences between behaviors of the fluid and gel states quantified by probing MSD at larger time scales.

**3D phototactic active nematics**

Without the specially designed 2D confinement discussed above, cyanobacterial filaments can freely move in 3D (Supplementary Movies 5 and 6). This enables the phototaxis-driven crowding of filaments while they move one atop another, with the ensuing formation of a 3D active nematic slabs with high concentrations (number densities) of trichomes (Fig. 3a-c). Simply shining light through a glass substrate leads to a gradual increase of the concentration of cyanobacteria within a volume above the illuminated region (Fig. 3a), within which the bacterial filaments can be relatively tightly packed to commonly span heights corresponding to 20-30 *Geitlerinema* sp. cell widths (Fig. 3b,c). The depth of phototaxis-driven bacterial community extension along the light propagation direction obtained in this way is, however, somewhat limited because of the source of visible light becoming eclipsed by the evolving filament assembly due to light absorption and scattering. To overcome constrains of similar scattering-related losses of light in the process of imaging, our experimental 3D spatial configurations of trichomes are reconstructed from stacks of depth-resolved fluorescence images acquired by multiphoton-absorption-based nonlinear optical fluorescence microscopy (see Methods and Fig. 3b-i), providing insights about the structure of the 3D samples with active nematic order. Cross-sectional images in planes parallel (Fig. 3d,e) and perpendicular (Fig. 3g-i) to **n** reveal the existence of fluidlike (Fig. 3d,f,g) and gellike (Fig. 3e,h) 3D active nematics, as well as the occasional formation of active fluid states with locally hexagonal near-field arrangements of geometric centers of filaments (Fig. 3i). Video-microscopy-based filament tracking in the lateral X-Y plane at 20 μm above the sample surface closer to the light source (Fig. 3j,k) reveals pronouncedly elongated filament trajectories in the lateral planes of the fluid state, which are oriented along the local director **n**. Similar to the case of 2D nematics, the active filamentous cyanobacterial particles here also exhibit ballistic propulsion along their longer axes parallel to **n** (with $\alpha_{\parallel} \approx 2$), whereas motility in directions perpendicular to **n** is hindered ($\alpha_{\perp} < 2$) (Fig. 3j-l), revealing strong anisotropy of the studied here cyanobacterial active nematics of both fluid and gel types. The analysis of MSD on the relatively short time scales (<100s) reveals that the cyanobacteria behave as active particles self-propelling along **n** in both nematic fluids and gels, though below we will reveal more complex behavior of MSD on larger time scales.

Similar to 2D nematics, the selection of director orientation in the growing monodomains of phototactic active nematic samples is an outcome of interactions between individual filaments, randomly chosen from different possible orientations, as we tested by using the same light source for samples repeatedly prepared under the same conditions. When the photactically induced domain is small, the average direction of cyanobacterial ordering slowly rotates over time, but such rotations stop as the domain grows to become large (hundreds of microns laterally).

The orientational order parameter of our 3D active nematic systems is calculated based on 3D nonlinear optical structural imaging and the velocity order parameter is estimated using a 2D nematic model[7] for both fluids and gels within the studied quasi-3D nematic states (see Methods). This approach of quantifying the 2D velocity order parameter in samples of modest thickness, typically smaller than filament lengths, is adopted because the cyanobacterial filaments tend to adopt horizontal quasi-2D orientations while exhibiting their motions in planes roughly parallel to the confining substrates. Both orientational and velocity order parameters exhibit relatively high values, within a typical range of 0.7-0.9, comparable to that of 2D nematic states, similarly obtained from videomicroscopy-based orientation distribution data shown in the Supplementary Fig. 3. Optical imaging with transmission-mode microscopy for thick samples (100µm and larger thickness) placed between crossed polarizers, with and without an additional retardation plate, reveals birefringence $\Delta n \sim 10^{-5}$ of the studied 3D phototactic nematic samples (see Methods and Supplementary Fig. 4). This finding suggests the presence of orientational order in organizations of not only the bacterial cells but also in the extracellular biopolymer molecules within the surrounding polysaccharide matrix, somewhat resembling observations for active nematics formed by cellulose-producing bacteria, where both bacteria and extracellular nanocellulose were found being orientationally ordered[23].

**Transformations between phototactic active nematic fluids and gels**

In addition to generating the birefringent medium of biopolymers co-assembled with trichomes, the polysaccharide slime (Supplementary Fig.1d) ejected by propelling cyanobacterial filaments eventually causes formation of a dense viscoelastic matrix[32,33]. As compared to isotropic-to-nematic transition, the 2D fluid-gel transformation stemming from this process happens continuously at longer time scales, typically over the course of days. This transition is accompanied by an additional gradual increase of filament number density as compared to the isotropic or nematic fluid states (Fig. 4a-c). For example, within 96 hours, starting from a relatively dilute cyanobacterial community with the surface coverage of $\sigma \approx 0.2$, a 2D nematic gel with $\sigma \approx 0.9$ and the corresponding orientational order parameter of $S_{2D} \approx 0.95$ (Fig. 4d) gradually forms. During this gel formation process, the amplitude of trichome velocity $V$ decreases down from the initial $V_0 = 2$ µm/s to values near zero (Fig. 4d), eventually making the system static, or "frozen." With the trichome velocity slowing down and bipolar motions featuring propulsion reversals at progressively shorter time scales, the MSD probed over $10^3$-$10^4$ seconds changes its behavior from being purely ballistic ($\alpha \approx 2$) in an isotropic low-density state to partially hindered in the nematic fluid and gel states (at relatively short time scales $\alpha \approx 1.6$), as revealed in Fig. 4e. With time and especially upon formation of nematic gels, the filament motility becomes quasiperiodic, which is manifested by the oscillating behavior of the longer-time-scale MSD curve (Fig. 4e, green curve, with the onset of such behavior also seen for the magenta curve describing the nematic fluid state). These changes in the MSD behavior at larger time scales are directly related to the transition from polar to bipolar motility with progressively shorter times and distances between bipolar motion reversals until a completely static state forms. Moreover, such polar-to-bipolar transformation in the motions behavior with elapsed time is found accompanying the isotropic-nematic and then fluid-gel transitions for samples both without (Figs. 2-5) and with (Figs. 6 and 7) topological defects, including both 2D and 3D nematic samples. On the time scales of tens of hours and days, the bacteria-secreted viscoelastic matrix creates tracks of cyanobacterial motions and, therefore, the propulsion character changes. In a nematic fluid, bacteria move roughly along **n** while following trajectories that evolve with time and feature relatively rare reversals of propulsion directions every several minutes (Fig. 5a). Cyanobacterial bipolar motions in the older samples within the gel state become more quasi-periodic in time, always along the same slime-defined tracks parallel to **n** (Fig. 5b). The typical distance $d_{rev}$ between the two neighboring motion reversal events of the same filament in a 2D active nematic gel is measured to be 22 µm (Fig. 5b,e), which is shorter than that in the corresponding fluid state (≥120 µm, Fig. 5a) and even the filament length itself (Supplementary Fig. 1d). Within this active fluid-gel transformation process, the average filament velocity drops from 0.8 µm/s in the nematic fluid state to 0.1 µm/s in the nematic gel state (Fig. 5f and Supplementary Movies 3 and 4). At the same time, both order parameters that we quantify, $S_{2D}$ and $\varphi$, feature high values in both the nematic fluids and gels, typically within the range of 0.6-0.9 (Figs. 2h,i and 4d).

The hindered filament dynamics in our active matter can be also probed by means of mapping the propulsion reversal points, as shown in Fig. 6a,b and Fig. 7a,b for fluid and gel states, respectively. As anticipated based on the above analysis, the reversal points are sparce in the 2D nematic fluid (Fig. 6a,b) as compared to the 2D nematic gel (Fig. 7a,b), differently from isotropic states with sparce filaments, where such propulsion reversals are rare (Supplementary Fig. 2a-c). The differences between our active nematic fluids and gels are also apparent from observing the recovery after photobleaching (Figs. 6c and 7c for the fluid and gel states, respectively), as further detailed in Supplementary Fig. 5. To obtain such experimental images and Movies, we photobleached a square-shaped region in both the fluid (Fig. 6c) and the gel (Fig. 7c) active matter media. The experiments reveal that the edge of a bleached region smears parallel to **n** in the fluid state, with recovery taking 12 min (Fig. 6c). In the gel state, differently, the edges of the bleached area appear sharper while the recovery takes ~40 min due to low trichome velocities (Fig. 7c).

As in 2D, the evolution of phototactic active matter behavior with time in our 3D active matter features a transition to quasi-periodic (oscillating) and confined spatiotemporal trajectories with much smaller displacements upon the formation of a gel state from the nematic fluid (Fig. 5c,d). The average $d_{rev}$ for these 3D nematic states is ~8 μm (Fig. 5e), interestingly, being even shorter than ~22 μm for its thin 2D counterpart of the same age. The filamentous velocity in the 3D nematic fluid, which is ~0.84 μm/s under illumination conditions of our experiments, eventually decreases down to 0.13 μm/s upon the contiguous formation of the gel state as time elapses (Fig. 5g). Photobleaching of both 3D fluids and gels reveals that the recovery to uniform states typically occurs within 30 minutes in the former case (Supplementary Fig. 5c) and only after more than 1 hour (Supplementary Fig. 5d) in the latter case, consistent with the observed slower filament propulsion within the gel revealed by video microscopy (Supplementary Movies 5 and 6).

**Emergence and behavior of defects in phototactic active nematics**

Trichome migration to lit areas often prompts merging of smaller nematiclike cyanobacterial colonies with misaligned local director orientations (Supplementary Figs. 6 and 7), which yields topological defects, singular points in the 2D **n(r)** where director cannot be defined and $S_{2D}$ vanishes[6] (Fig. 6d-j, Fig. 7d-i and Supplementary Movies 7 and 8). We observed defect formation for both *Geitlerinema* (Fig. 6d-j, Fig. 7d-i and Supplementary Fig. 6) and *O. brevis* (Supplementary Fig. 7) species studied here. This defect formation out of the misaligned nematic domains can be thought of as an active matter analog of the Kibble-Zurek mechanism[35-37], where nematic domains having different director orientations merge to form the topological defects (Supplementary Fig. 6). The probabilities of having defects or larger uniform domains formed in this process depends on the misalignment angles between multiple interacting smaller domains. The observed defects are characterized by a winding number *m*, defined as the number of times the director rotates by $2\pi$ when one circumnavigates the defect once, with the sign determined by the **n(r)** rotation relative to the circumnavigation direction. Our phototactic nematics exhibit 2D defects of $m = +1/2$ (Fig. 6d-f and Fig. 7d-f) and $m = -1/2$ (Fig. 6g-i, and Fig. 7g-i), consistent with both the nonpolar symmetry of **n(r)** and the bipolar nematic cyanobacterial motility. Described by mappings from a circle ($\mathbb{S}^1$) surrounding the 2D nematic's defects to the $\mathbb{S}^1/\mathbb{Z}_2 \equiv \mathbb{S}^1$ order parameter space, these $\pi_1(\mathbb{S}^1/\mathbb{Z}_2)=\mathbb{Z}$ first homotopy group's elementary defects are the only ones observed experimentally in our 2D nematics, even though the higher-winding-number defects with $m=\mathbb{Z}/2$ (as defined here following conventions[38]) are topologically allowed. This is likely because of higher energetic costs of their high-$m$ counterparts in terms of orientational elasticity[6,38]. The +1/2 defects in the nematic fluid state (Fig. 6d-f and Supplementary Movie 7) exhibit propulsion, a feature common for the active matter systems where such defects themselves are described as polar active quasi-particles[1,16]. For conventional active nematics, similar propulsion of +1/2 defects with the "comet head" forward would reveal presence of flows with extensile active stresses.[16] In our system, active-particle-like defect motility also results from the lack of balance of active forces within the 2D active nematic host around the singular core of the +1/2 defect, which can be also linked to the asymmetry in density of points of propulsion reversals revealed by videomicroscopy tracking (Fig. 6f), where extra reversal points result from misaligned trichomes bumping into one another at the defect core's site and, thus, propelling the defect. Differently, the −1/2 defects behave like passive quasi-particles, for which such active forces are balanced due to higher inherent symmetry (Fig. 6g-i and Supplementary Movie 7).[16] Spatiotemporal evolution of $\mathbf{n}(\mathbf{r})$ around defects in nematic fluid state is shown in Fig. 6f and i, where local average filament orientations are depicted before and after the temporal evolution over a short fragment of time, along with the points where the trichomes reverse their propulsion directions. The reversal points densely populate the +1/2 defect core's path (Fig. 6e-f), consistent with the discontinuity of $\mathbf{n}(\mathbf{r})$ around these asymmetric singular points that require the reversal to occur and that provide further insights into physical origins of the +1/2 defect motions. Annihilation of defects with opposite $m$ (Fig. 6j and Supplementary Movie 7) invokes primarily the motion of a +1/2 defect towards its −1/2 counterpart, similar to the case of other active nematics.[14-16] Since the typical initial defect density is low, around 2-4 defects per 1 mm$^2$, such annihilation events are rare and tend to drive the system towards a monodomainlike nematic state. Unlike in other active nematic systems[14-16], where defects were often found both annihilating and being generated in pairs, events of activity-driven generation of defects with opposite $m$ so far have not been observed in the cyanobacterial phototactic active matter system under conditions of our experiments, which was anticipated for active nematics in a certain parameter range for which activity-related forces cannot overcome the orientational-elasticity-mediated attraction between the defects.[16] In our system the phototactic effects also tend to promote annihilation because defect cores are deprived of trichomes. Thus, the monodomain regions tend to localize in the illuminated areas and defects tend to be expelled outside of them while often also annihilating. While theoretical foundations of such effects will need to be considered in the future, this simple explanation is consistent with the observation that relatively small, illuminated regions of phototactic active nematics tend to be defect-free in their central parts.

Continuous polysaccharide slime ejection by the cyanobacteria while they move[33] (Supplementary Fig. 1d), as well as the eventual formation of a dense viscoelastic matrix in the ensuing dense bacterial communities[32], strongly hinder the defect dynamics (Fig. 7). Consequently, while the presence of motile +1/2 defects confirms the initial fluidlike 2D phototactic nematic state (Fig. 6d-f), a transition from fluid into gel-like state occurs over time due to slime accumulation and immobilizes the +1/2 defects too (Fig. 7d-i). This transformation is accompanied by an increase of the number of the filament propulsion reversal points. Topological defects in the well-developed nematic gel states are immobile regardless of their charge or symmetry (Fig. 7d-i and Supplementary Movie 8), featuring trichomes that propel along their static "frozen" tracks decorating (following) the $\mathbf{n}(\mathbf{r})$ (Fig. 7f and i). Like for filaments themselves, the velocity of +1/2 defects reduces continuously down to zero during the fluid-gel transformation. The semi-flexible nature of filaments allows them to follow highly curved trajectories while moving around the defect cores (Fig. 7e and h), without getting off the slime-created tracks. Velocity values and profiles of filaments moving around defect cores in the nematic fluid (Fig. 8a,b) and gel states (Fig. 8c,d) reflect both the symmetry of the director field and the nematic fluid-versus-gel state. A comparison of velocity profiles (Fig. 8a-d) and corresponding probability distributions (Fig. 8e,f) can be used to quantify the intrinsic fluidity of the active nematic medium. For example, the average trichome velocity around defects in a nematic gel is 0.1-0.2 μm/s, which is lower than that in the nematic fluid state (within 0.5-0.6 μm/s, Fig. 8e-f), and even lower than the typical velocity of ~0.3 μm/s of the +1/2 defect's core moving within the 2D phototactic nematic fluid.

At even larger spatial scales (tens-to-hundreds of microns in thickness), 3D phototactic active nematics can exhibit complex topological defects (Fig. 9a and Supplementary Movie 9), with the filament motion trajectories and director field $\mathbf{n}(\mathbf{r})$ tilting out of the planes parallel to confining substrates and becoming 3D in nature, with various disclination lines propagating through the nematic bulk. Belonging to the 1st homotopy group $\pi_1(\mathbb{S}^2/\mathbb{Z}_2)=\mathbb{Z}_2$ (Fig. 9b-d and h),[38] these disclinations can have their local structures morphed between geometrically different states and feature only one type of topologically stable objects in 3D that differs from the topologically trivial state and never terminates within a nematic bulk. Although their local cross-sectional structures resemble those of +1/2 and −1/2 pointlike defects in the 2D phototactic nematics (Fig.

9d-k), these structures can smoothly morph one to another in 3D while meandering through the active nematic bulk. 3D cross-sectional nonlinear optical imaging allows for deep noninvasive penetration of optical scanning (enabled by near-infrared femtosecond excitation light) and robustly elucidates the **n(r)** of these disclination lines (Fig. 9e-g and i-k).

Hypothetically, topology of 3D active nematics also allows for hosting $\pi_2(\mathbb{S}^2/\mathbb{Z}_2)=\mathbb{Z}$ point defects in the nematic bulk,[38] elements of the 2nd homotopy group, but so far they could not be found as bulk defects in our experiments. This might be due to energetic or kinetics of formation related reasons that for our samples could be partly due to relatively small thickness of samples. Coincidentally, this lack of observation of topologically allowed bulk point defects is also similar to recent findings for other types of 3D active nematic systems.[15] However, the reasons for the lack of observation of point defects here might be different and remain to be elucidated as, for example, the phototactic expulsion of defects with cores deprived of trichomes out of illuminated volumes (similar to expulsion of 2D defects in 2D nematics discussed above) may play a role.

**Passive inclusions and lateral confinement with light**

Surface confinement prompted by formation of droplets or via immersing colloidal particle inclusions is often used to induce defects in nematic systems, even under circumstances when without confinement such defects quickly annihilate in the passive nematic liquid crystal's bulk.[38,39] Furthermore, for strong tangential or perpendicular surface boundary conditions the topological invariants of defects can be related to the topological characteristics of confining surfaces, like Euler characteristics or genus, including cases of spheres, handlebodies and other geometric structures and topologies.[39-43] In our phototactic cyanobacterial active nematics, tactoid-like active nematic droplets can be formed using strongly localized light shined into sample areas of interest. For example, Figure 10a-f reveals formation of an active nematic state roughly within the area of an illuminated square-shaped region, with the surrounding areas practically deprived of cyanobacterial filaments. As the filaments move into this square-shaped illuminated area, they eventually adopt a nematic state that has the director mainly oriented along a diagonal of the square (Fig. 10a). This along-diagonal direction of the cyanobacterial filament orientations and motions emerges spontaneously (alignment along the other diagonal can occur with the same probability). Sharp corners of the square impose weak perturbations of the alignment and motion directionalities (bottom left and top right corners in Fig. 10a) or cause locally reduced number density of bacteria (top left and bottom right corners, Fig. 10a). The cyanobacterial filament orientations slightly depart from the orientation in the bulk along the other fragments of the perimeter of the illuminated square-shaped region (Fig. 10a-f). In the terminology common for conventional thermotropic liquid crystals, this observation corresponds to "weak tangential surface boundary conditions", where near-edge filaments tilt away from that in bulk in a way that reduces the angle between the square-shaped perimeter and cyanobacterial filaments (Fig. 10a-f). The top right and bottom left corners of the light-confined active nematic region feature the surface counterparts of bulk point defects, called boojums.[38-43] Like in the examples featured in Fig. 3, the tactoidal active nematic droplet induced by light has thickness larger than that of an individual filament. Although the misalignment of director and velocity directionalities allows for assigning fractional charges of the ensuing boojum-like defects,[38] the geometric features of these defects are rather different from what one finds in passive nematic droplets of nanometer sized rodlike molecules, though they appear to be somewhat closer to the counterparts of such defects in colloidal systems.[39-43]

Boojum-like defect features[38-43] were also observed being induced at the surfaces of foreign inclusions in the form of salt crystals (Fig. 10g-j and Supplementary Movie 10) of different geometric shapes, which are often found embedded into monodomain active nematics after partial evaporation of the nutrition medium hosting the cyanobacterial communities. Such quasi-boojum features are especially vividly seen for larger inclusions, as marked in the bottom parts of Fig. 10i,j. We note that the typical lengths of the studied cyanobacterial filaments are larger than these immobile surface-attached foreign inclusions, a situation once again very different from various inclusions in passive nematic systems where nanometer-long molecules are much smaller than micrometer-sized particles for which boojums are often observed forming at the poles along the far-field director.[38-43] Quasi-boojums also occur in conventional passive nematic colloidal systems under equilibrium conditions, where the lengths of rodlike particles can be also comparable to inclusion diameters, contributing to large cores of these defects,[38-43] similar to how such quasi-boojum cores are deprived of cyanobacteria in our system (Fig. 10g-j and Supplementary Movie 10). In future studies, it will be of interest to explore how various mobile and immobile inclusions with different surface topology and geometric shapes interact with cyanobacterial active nematics in the regimes of inclusions and confining features being larger than the filament lengths. This exploration may be of especial interest from the standpoint of inducing various topological defects and controlling their dynamics by exploiting confinement geometry and viscoelastic effects[44,45] that can be synergistically combined with the phototactic effects introduced in this work.

**Discussion**

The demonstrated light-induced liquid crystallinity of active matter is an effect distantly analogous to phototropism of conventional passive nematic systems formed by photosensitive molecules.[46] The studied cyanobacterial filaments, powered by a photosynthetic conversion of light into energy,[47] readily self-organize into large contiguous domains of 2D and 3D active nematics upon locally shining unstructured light with spatial intensity gradients. In this process, the motility of individual semiflexible cyanobacterial filaments changes from being polar and towards the light or along complex curved trajectories within illuminated areas of initially sparce communities of filaments to becoming bipolar and along straight lines once the nematic state forms. While models considering concentration and collisions of rigid rodlike active particles in 2D[5,7,8] anticipated the formation of active nematic with increasing their concentration, here this emergent process in both 2D and 3D is much more complex than what was previously modeled or experimentally observed (Figs. 2 and 3). This is because the phototaxis-mediated crowding of cyanobacterial filaments also emergently transforms their motility from being mainly polar and along complex curved trajectories localizing bacterial motions within illuminated areas to bipolar straight-line dynamics (Supplementary Figs. 1-3 and Movies 2-4). From a biological perspective, the trichomes' transition to bipolar motility within the crowded environment aids in regulating intake of light-supplied energy by allowing the cells to visit the area of illumination repeatedly, thus securing the access to the energy source. Furthermore, the crowding and spontaneous ordering is further accompanied by increasing the average length of trichomes and the slime secretion, eventually yielding dense 2D packing of long filaments with polydisperse lengths (Figs. 2 and 3 and Supplementary Movies 3 and 4). Moreover, contrary to naïve expectations, such effects could not be reproduced upon simply concentrating the cyanobacterial filaments by centrifugation (upon which cell viability is compromised and the centrifuged cells die, likely due to the lack of slime lubrication between them), highlighting the importance of our methodology of obtaining phototactic active nematics that is compatible with natural biological processes. Furthermore, the temporal evolution of order parameters depends on the intensity of light illumination (Supplementary Fig. 8): the dimmer light, the slower nematic emergence, with lower values of the orientational order parameter S during evolution and in the developed over long time fluid/gel states; no nematic states form at very low intensities (0.01 cd and lower). Thus, our findings reveal rather complex phototactic-behavior-related origins of the large-scale active nematic formation, showing behavior markedly different from that found under uniform illumination conditions without light intensity gradients, where orientational order develops locally in the form of interconnected active-spaghetti-like domains.[49] New theories describing emergence of nematic order under phototaxis-driven transition will need to be developed. Furthermore, we have also revealed how nematic fluids without and with motile defects eventually transform into gels without and with static defects, also featuring quasi-periodic oscillating motions of filaments along tracks decorating the nematic director field. These phototactic nematic fluids and gels are new breeds of active matter that hold a great potential for expanding its fundamental and technological utility.

Our experimentally highly accessible system may offer unprecedented means of controlling active matter by supplying energy via patterned illumination, as well as may allow to model diverse out-of-equilibrium effects. For example, while recent experimental, numerical and theoretical studies already pursue control of activity and defect dynamics with light,[48,49] our system offers different means of exerting such control due to the combination of its light-powered photosynthetic and phototactic nature that was not utilized prior to this our work. While previous works exploit light sensitivity of already existing nematics by locally controlling activity, our study opens doors to on-demand creation of 2D or 3D active nematics in desired areas or volumes of arbitrary shape or topology, as well modulating activity to adopt arbitrary spatiotemporal patterns. For example, our 3D multi-photon-absorption based imaging of 3D phototactic active nematics (see Methods) utilizes the very same nonlinear optical effects that can be also used to induce nematic states and photosynthetically activate cyanobacterial motility along arbitrary patterns, like the ones corresponding to 3D topologies of trefoil and other knots, Hopf and other links, etc.[38,39] While prior studies explored how defects in active nematics behave under various conditions, our work reveals how small domains of cyanobacterial active nematic with different orientations can can come together to give origins to defects, an active matter analog of the celebrated Kibble-Zurek defect formation mechanism. One can imagine, for example, designing conditions to probe statistical evolution of defects as they evolve upon the contiguous nematic formation from smaller nematic domains, analogously to studies of the Kibble-Zurek mechanism in passive condensed matter systems. However, realizing and exploiting the entire spectrum of possibilities enabled by our work is well beyond the scope of this present study and will be pursued elsewhere, as well as we anticipate will open new active matter research directions.

Our study can be extended to a large variety of cyanobacteria and other light-responsive lifeforms, opening the doors to optically reconfigurable and stimuli-responsive active matter with unprecedented control of physical behavior.[50] While our active phototactic nematics were found to spontaneously form for the two species studied, the vast variety of cyanobacterial morphologies (e.g. shaped as spirals, branched filaments, etc.) may allow for the formation of more complex active liquid crystals, like active cholesterics, as well as may allow for exploring the inter-relations between diversity of individual bacterial motility (e.g. spinning while translating, or not) and that of the ensuing self-assembled active matter. Thus, facile responses of these active particles to light and light intensity gradients via phototaxis may allow their use as a model system in realizing new forms of orientationally ordered active matter. From a more applied perspective, gradients of intensity could be generated and dynamically reconfigured to maximize or inhibit many processes that are important in the context of natural cyanobacterial habitats, ranging from producing oxygen, biofuels, biomaterials or inhibiting unwanted production of toxins.[24-30]

## Methods

**Bacterial cultivation**

Our experiments were performed on two cyanobacterial species: the motile filamentous cyanobacterium *Geitlerinema* sp., which was kindly provided by Dr. Elizabeth Trower from the University of Colorado Boulder (Department of Geological Sciences), and the model of *Oscillatoria brevis* (*O. brevis*) purchased from the UTEX, Culture Collection of Algae at University of Texas at Austin (UTEX B 1567, isolated from brackish pond water collected in Miami, Florida, USA). Both species live colonially in long filaments, also called trichomes, consisting of tens to hundreds of cells (Supplementary Fig.1). The *Geitlerinema* sp. cells possess rodlike shape with average width of 2.5 µm and length of 3.7 µm, whereas the *O. brevis* cells are more disklike in shape with dimensions of 3.7 µm and 1.8 µm, respectively (Supplementary Fig. 1). We used A+ nutrient medium to cultivate *Geitlerinema* sp., and BG-11 medium to cultivate *O. brevis*. A+ was prepared following the standard protocol[51], whereas BG-11 was purchased from UTEX. Both species were initially grown on top of solidified agar set in the corresponding nutrient medium at a density of 1% w/v. The growth took place over 5 days in petri dishes placed in an incubator (Innova 4080, New Brunswick Scientific) with the temperature fixed at 30°C, under continuous illumination with saturating light level of ~70 µmol photons $m^{-2}$ $s^{-1}$ (~200 cd) provided by a cool fluorescent lamp (Phillips). Afterwards, for further exponential growth, cells were transferred to 25 ml of A+ and BG-11 liquid media as appropriate to the species in 125 ml baffled flasks capped with a foam stopper (Thermo-Fisher). Growth in liquid medium took place over 5-7 days in the incubator under orbital shaking (120 r.p.m.) with fixed temperature (30°C) and illumination (~70 µmol photons m$^{-2}$ s$^{-1}$ corresponding to ~200 cd). The typical *Geitlerinema* sp. and *O. brevis* filament length in freshly grown colonies dispersed in the liquid medium was 200-400 µm and 500-700 µm respectively (Supplementary Fig. 1). In our experiments, we additionally used sample preparations with the initially shorter yet polydisperse filaments of typical length ranging from 20 µm to 100 µm. The length shortening was achieved by gentle filament sonication over 20-40 seconds in a ultrasonic bath (purchased from Bransonic) or by a procedure of "aging" the culture for approximately one month. This "Aging" procedure is implemented by transferring culture dishes from incubator conditions to a benchtop and growing them under ambient light (diurnal light cycle of sunlight at ~12 µmol photons m$^{-2}$ s$^{-1}$) at 20°C.

**3D and 2D phototactic active nematic sample preparation**

To obtain the 3D nematic cyanobacterial communities, we initially placed around 7 ml of filaments diluted in their nutrition media at typical concentration of 10$^3$-10$^4$ cells µl$^{-1}$ into an imaging dish (µ-dish, purchased from Ibidi) with 35mm or 50mm diameter and #1.5H, 170 µm thickness glass coverslip bottom. This sample configuration allows for optical microscopy observation with high-numerical-aperture and relatively short-working-distance objectives, as needed for high-resolution optical imaging. To achieve geometrical 2D confinement (Fig. 1), bacterial communities were additionally sandwiched between the µ-dish bottom surface and the uniformly flat solidified agar medium (slab area ~2x2 cm$^2$) infused with the nutrients. The initially low cell concentration allowed for uniform one-bacterium-thick spacing distance between the dish bottom and agar surface, which in turn is favorable for the cyanobacterial filament monolayer formation (Fig. 2e).

**Application of localized illumination**

Depending on the aimed dimensions of both 2D and 3D nematic bacterial communities, we applied different types of photo-patterning to the filament dispersions in cell culture imaging µ-dishes at either submillimeter (Fig. 2a-b) or centimeter scales (Fig. 3a, Fig 9a and Fig. 10a). At the submillimeter scale, we utilized light sources of an Olympus IX-83 inverted microscope, which can function in either brightfield illumination mode or confocal laser scanning mode. The µ-dish with bacteria dispersions was positioned onto the microscope's motorized stage equipped with a specially designed sample holder. The source of white light directed from the top was from a light source of brightfield microscopy imaging mode, a halogen bulb. Another light source was the λ=561nm (Coherent OBIS) monochromatic laser beam directed from the bottom, normally used as excitation source in the confocal microscopy imaging mode. Both light sources could be used for local illumination of the studied samples, which phototactically attracted and confined the trichomes in the lit area (Fig. 1a-c). The size of the illuminated area was controlled by a condenser aperture diaphragm in the case of white-light illumination, and with the help capabilities of the FluoView FV3000 software (Olympus) that could define illuminated areas in the laser scanning mode. All experiments were conducted in near-total-dark environments (photosynthetic photon flux corresponding to ~0.1 µmol photons m$^{-2}$ s$^{-1}$). The local illumination light in both approaches described above produced photosynthetic photon flux of ~2-3 µmol photons m$^{-2}$ s$^{-1}$ (corresponding to ~1.5-2 cd). Cell division in µ-dishes was negligible under our experimental conditions due to lack of the light energy input. Moreover, the isotropic-to-nematic transitions and other effects could be directly observed and characterized with single-cell resolution using an Olympus IX-83 microscope and its illumination controls for both imaging and photo-patterning.

To implement the phototactic optical control of the cyanobacteria at the centimeter scales (Fig. 3a, Fig 9a and Fig. 10a), we placed the µ-dishes with bacteria dispersions atop of a horizontally oriented Dell liquid crystal display (1920 x 1080 pixels resolution) with computer-programmed illumination. The desired patterns of light intensity and color were then defined by arrays of pixels of the monitor with desired light illuminations of red, green, white or other color applied to the samples over illumination areas of arbitrary shapes, with and without gradients in intensity, and at millimeters-to-inches lateral dimensions. The experiments were conducted in a dark room under negligibly low ambient light levels of no more than ~0.1 µmol photons m$^{-2}$ s$^{-1}$. At the same time, the pre-designed locally illuminated cyanobacterial community areas above the Dell display were continuously exposed to ~1 µmol photons m$^{-2}$ s$^{-1}$ (~ 0.07 cd) to phototactically control filaments. Glycerol was sandwiched between the monitor and the µ-dish bottom to reduce light reflections and scattering from the surfaces, effectively serving as an index-matching medium. To examine the filaments' assembly and isotropic to nematic phase transition in these display-enabled patterned illumination experiments, the µ-dish was removed and observed using the optical microscope described below, acquiring images and data on structure and dynamics of the cyanobacterial active matter.

The active nematic state of motile *Geitlerinema* sp. filaments at the water-air interface (Fig. 2d) was achieved by bacterial cultivation in an incubator in a µ-dish sealed to the environment under static conditions at fixed temperature (30°C) and for overall illumination yielding a photosynthetic photon flux of ~70 μmol photons m$^{-2}$ s$^{-1}$ (or ~200 cd). The growth of salt crystals within the bacterial communities (Fig.10g-j) was achieved by controlled A+ medium evaporation over 2-3 days at room temperature (20°C) and ambient light (diurnal light cycle of sunlight at ~12 μmol photons m$^{-2}$ s$^{-1}$).

**Optical microscopy observations and 3D nonlinear optical imaging**

To study cyanobacterial communities, we have utilized an Olympus IX-83 inverted microscope in the brightfield imaging mode integrated with FV-3000 confocal fluorescence imaging system, which was controlled by FluoView software (from Olympus). Conventional white light (halogen bulb) was employed for brightfield imaging. Monochromatic laser beams with excitation wavelengths of $\lambda_{ex}$ = 405nm targeting chlorophyll or $\lambda_{ex}$ = 561nm targeting phycobilin (phycocyanin/phycoerythrin) were used as light sources for fluorescent imaging. These beams were generated by the corresponding Coherent OBIS Cube lasers with maximum output power of 50 mW, though only 0.1-2% of this power was used to assure minimally destructive imaging. All cyanobacteria are endogenously fluorescent due to the presence of light-capturing chlorophyll and other pigments such as phycobilins[51]. The detected emission wavelength ranges in our experiments were 660-760nm for chlorophyll and 630-730nm for phycobilins. To allow for high-resolution fluorescent imaging of 3D dense and thick nematic domains while reducing light scattering (Figs. 3 and 9), two-photon absorption nonlinear optical fluorescence microscopy was utilized. In this mode, a Coherent Chameleon Discovery tunable femtosecond laser was used to generate excitation beams at wavelengths of $\lambda_{ex}$ = 810nm and $\lambda_{ex}$ = 1122nm for chlorophyll and phycobilin, respectively. The detected emission wavelength ranges were the same as for single-photon confocal fluorescence imaging mode described above. In some cases, to enhance the amount of collected fluorescent light emitted from the cyanobacterial samples, a forward-detection mode was used by using an additional objective lens (50x air objective, NA=0.5), an external photomultiplier tube (PMT) detector, and a series of optical filters. The fluorescence signal collected by the objective lens was separated from stimulating light with a long-pass dichroic mirror for only diverting the fluorescent light to the PMT detector and a short-pass filter (775 nm edge wavelength and T$_{avg}$ > 93% 481-756 nm) for further rejecting the excitation light. A band-pass filter before the detector was also used for further cleanup of detected signals.

Structural information characterizing the trichomes' orientational order at large scales (Figs. 3a, 9a and 10a) was acquired using a dry 10x (NA=0.42) objective. Dynamical properties were mainly probed by using optical microscopy with oil immersion 60x or 100x objectives (NA=1.4) to allow for high-resolution imaging, as needed for proper filament motility tracking (as in examples shown in Fig. 2c). To acquire 3D image stacks with multiphoton absorption nonlinear optical fluorescence microscopy, the 60x (NA=1.4) oil immersion objective was used.

Videos with cyanobacteria in different active matter states were recorded over typical 30-120 min time windows. The stability of the microscope's stage over these time windows was achieved by the use of a Z-drift compensation system. For brightfield observations framerates of 0.2-1 fps (Hz) were used during recording. In confocal fluorescent and multiphoton nonlinear optical imaging modes, scanning speeds of 2-5 seconds per slice (0.2-0.5 Hz) were used. The cross-sectional optical slice's lateral scanning dimensions ranged from 200x200 µm² to 1200x1200 µm², depending on the objective magnification. The typical saturating level of light used for imaging was set as low as ~0.3-0.5 µmol photons m$^{-2}$ s$^{-1}$ (~0.2-0.4 cd) both to reduce the effect of imaging light on phototactic motility and to minimize cyanobacterial photobleaching. For the photobleaching recovery experiments, square-shaped regions of 160x160 µm² were photobleached extensively using monochromatic laser light at $\lambda_{ex}$ = 405nm or $\lambda_{ex}$ = 561nm with the saturating light level of ~20-100 µmol photons m$^{-2}$ s$^{-1}$ (30-80 cd) over 10-30 min.

To reveal and characterize birefringence in our nematic samples (Supplementary Fig. 4), an Olympus BX-51 polarizing optical microscope was used. The additional crossed linear polarizers and a 530nm full-wave phase retardation plate, needed to enable the polarizing optical microscopy mode observations, were utilized in these observations. The optical birefringence, defined as $\Delta n = n_e - n_o$, where $n_e$ and $n_o$ are the extraordinary and ordinary refractive indices of the optically monocrystalline uniaxial nematic medium, was measured by using a Berek compensator[52] for the bacterial communities developed in the µ-dishes. The nematic director **n** in our measurements was first oriented at 45° with respect to crossed polarizers to maximize light transmission and then rotated to confirm the monodomain nature of samples via continuous rotation between the crossed polarizers, followed by a periodic variation of light transmission.

**Characterization of order and motility of cyanobacterial filaments**

To analyze cyanobacterial filament surface coverage, orientation and motility, these active particles were tracked using an open-source ImageJ/Fiji "Manual Cell Tracking"[53] or, whenever possible, with an automatic "wrMTrck" plugin[54] (freeware from NIH). The cyanobacterial 2D relative surface coverage area $\sigma$ was measured using low magnification optical microphotographs (1200x1200 $\mu m^2$) and the definition $\sigma=A/A_0$, where $A$ is the area of the observed region, and $A_0$ is the area that is covered by the filaments. The error bar for the measured $\sigma$ at a given time $t$ (Fig. 2g) was calculated as a standard deviation $(x^2/k-1)^{1/2}$, where $x^2$ is the square of the surface coverage deviations on each image from the mean value, and $k$ typically equals 5 and is the number of consecutive images taken at moments $t-2\Delta t, t-\Delta t,...,t+2\Delta t$ ($\Delta t$ is time step between the two consecutive images).

The orientational order parameters were calculated through the numerical integration of the filament angular distributions using the expression[6] $S_{2D}=<2cos^2\theta-1>$ known for 2D nematic systems and $S_{3D}=1/2<3cos^2\theta-1>$ for the 3D systems, respectively, where $\theta$ is the angle between the individual trichomes and the local director **n**. Error bars for $S_{2D}$ and $S_{3D}$ were calculated as a standard deviation based on analyzing 5 consecutive fluorescent images (similar to analyzing errors for the surface coverage $\sigma$). Filament velocity distributions, $d_{rev}$, mean square displacement (MSD), mean square angular displacement (MSAD), and velocity order parameter $\varphi$ were calculated while employing MATLAB-based codes and relying on typically 30 filament trajectories extracted from the video-microscopy data. The typical temporal trajectory lengths were ~100 frames in fluid states and ~300 frames for gel states. The cyanobacterial velocity $V$ at a given time $t$ was calculated as $V=|\mathbf{r}_t - \mathbf{r}_{t-1}|/\Delta t$, where $\mathbf{r}_t$ and $\mathbf{r}_{t-1}$ are the spatial positions of filaments in the two consecutive frames. The parallel and perpendicular MSDs of bacteria under the nematic fluid or gel formation conditions were quantified with the expression $MSD = (N-n)^{-1}\Sigma_i^{N-n}(\mathbf{r}_{i+n}-\mathbf{r}_i)^2$, where $n$ is the number of intervals of a particular lag time (e.g. the time elapsed between consecutive frames of the video), $N$ is the total number of lag times, and $\mathbf{r}_i$ is the parallel position of a given cyanobacterium, respectively. Analogously, the MSAD values were calculated[55] using the angle values $\theta$ at each frame $MSAD = (N-n)^{-1}\Sigma_i^{N-n}(\theta_{i+n}-\theta_i)^2$. The velocity order parameter $\varphi$ was computed for 2D systems using the expression for the 2D nematic velocity order parameter $\varphi = <\varphi(t)>_t = 2((<cos^2\chi>_t - 1/2)^2 + <sin\chi cos\chi>^2_t)^{1/2}$, where $\chi$ is the angle describing orientation of the velocity vector relative to **n**, measured by analyzing the video-microscopy frames[7]. The velocity order parameters in quasi-3D active nematic systems were calculated in a way similar to quantifying $\varphi$ in 2D phototactic active nematics described above, which is because the active nematic samples of modest thickness feature distributions of velocity vectors effectively confined to two-dimensional planes parallel to the confining substrate.

**Data availability:** The data generated in this study are provided in the Source Data file. All other data that support the plots produced in this paper, as well as other findings reported in this study, are available from the corresponding author upon a reasonable request.

**Code availability:** The codes used for the numerical calculations are available upon request.

**References**

1. Marchetti, M. C. et al. Hydrodynamics of soft active matter. *Rev. Mod. Phys.* **85**, 1143-1189 (2013).

2. Ramaswamy, S. The Mechanics and Statistics of Active Matter. *Annu. Rev. Conden. Ma. P.* **1**, 323 (2010).

3. Toner, J. & Tu Y., Flocks, herds, and schools: A quantitative theory of flocking. *Phys. Rev. E* **58**, 4828 (1998).

4. Vicsek, T. et al. Novel type of phase transition in a system of self-driven particles. *Phys. Rev. Lett.* **75**, 1226 (1995).

5. Vicsek, T. & Zafeiris, A. Collective motion. *Phys. Rep.* **517**, 71-140 (2012).

6. Chaikin, P. M., Lubensky, T.C. & Witten, T.A. Principles of Condensed Matter Physics (Cambridge: Cambridge University Press,1995).

7. Chaté, H., Ginelli, F. & Montagne, R., Simple model for active nematics: Quasi-long-range order and giant fluctuations. *Phys. Rev. Lett.* **96**, 180602 (2006).

8. Ginelli, F., Peruani, F., Bär, M. & Chaté, H. Large-scale collective properties of self-propelled rods. *Phys. Rev. Lett.* **104**, 184502 (2010).

9. Mushenheim, P.C. et al. Dynamic self-assembly of motile bacteria in liquid crystals. *Soft Matter* **10**, 88-95 (2014).

10. Smalyukh, I.I. et al. Elasticity-mediated nematiclike bacterial organization in model extracellular DNA matrix. *Phys. Rev. E* **78**, 030701 (2008).

11. Peng, C. et al. Command of active matter by topological defects and patterns. *Science* **354**, 882-885 (2016).

12. Saw, T.B. et al. Topological defects in epithelia govern cell death and extrusion. *Nature* **544**, 212-216 (2017).

13. Balasubramaniam, L. et al. Investigating the nature of active forces in tissues reveals how contractile cells can form extensile monolayers. *Nat. Mater.* **20**, 1156-1166 (2021).

14. Sanchez, T. et al. Spontaneous motion in hierarchically assembled active matter. *Nature* **491**, 431-434 (2012).

15. Duclos, G. et al. Topological structure and dynamics of three-dimensional active nematics. *Science* **367**, 1120-1124 (2020).

16. Giomi, L., Bowick, M. J., Ma, X. & Marchetti, M. C. Defect annihilation and proliferation in active nematics. *Phys. Rev. Lett.* **110**, 228101 (2013).

17. Sumino, Y. et al. Large-scale vortex lattice emerging from collectively moving microtubules. *Nature* **483**, 448-452 (2012).

18. Shankar, S. et al. Topological active matter. *Nat Rev Phys* **4**, 380-398 (2022).

19. Palacci, J. et al. Living crystals of light-activated colloidal surfers. *Science* **339**, 936-940 (2013).

20. Bililign, E. S. Motile dislocations knead odd crystals into whorls. *Nat. Phys.* **18**, 212-218 (2022).

21. Brandenbourger, M., Locsin, X., Lerner, E. & Coulais, C. Non-reciprocal robotic metamaterials. *Nat. Commun.* **10**, 1-8 (2019).

22. Bain, N. & Bartolo, D. Dynamic response and hydrodynamics of polarized crowds. *Science* **363**, 46-49 (2019).

23. Repula, A., Abraham, E., Cherpak, V. & Smalyukh, I. I. Biotropic liquid crystal phase transformations in cellulose-producing bacterial communities. *Proc. Natl. Acad. Sci. USA* **119**, e2200930119 (2022).

24. Kasting, J. F. & Siefert, J. L. Life and the evolution of Earth's atmosphere. *Science* **296**, 1066-1068 (2002).

25. Biddanda, B. A. et al. Seeking sunlight: rapid phototactic motility of filamentous mat-forming cyanobacteria optimize photosynthesis and enhance carbon burial in Lake Huron's submerged sinkholes. *Front. Microbiol.* **6**, 930 (2015).

26. Kamennaya, N. A. et al. High pCO2-induced exopolysaccharide-rich ballasted aggregates of planktonic cyanobacteria could explain Paleoproterozoic carbon burial. *Nat. Commun.* **9**, 1-8 (2018).

27. Mullineaux, C. W. & Wilde, A. Bacterial Blooms: The social life of cyanobacteria. *Elife* **10**, e70327 (2021).

28. Huisman, J. et al. Cyanobacterial blooms. *Nat. Rev. Microbiol.* **16**, 471-483 (2018).

29. Maeda, K. et al. Biosynthesis of a sulfated exopolysaccharide, synechan, and bloom formation in the model cyanobacterium Synechocystis sp. strain PCC 6803. *Elife* **10**, 1-19 (2021).

30. Moore, K.A. et al. Mechanical regulation of photosynthesis in cyanobacteria. *Nat. Microbiol.* **5**, 757-767 (2020).

31. Leeuwenhoek, A. V. More observations from Mr. Leeuwenhoek, in a letter of Sept. 7. 1674. sent to the publisher. *Philosophical Transactions of the Royal Society of London* **9**, 178-182 (1674).

32. Okajima, M. K. et al. Cyanobacteria that produce megamolecules with efficient self-orientations. *Macromolecules* **42**, 3057-3062 (2009).

33. Adams, D.G. How do cyanobacteria glide? *Microbiology Today* **28**, 131-133 (2001).

34. Stevens MK Faluweki, L Goehring. Structural mechanics of filamentous cyanobacteria. J. Royal Society Interface **19**, 20220268 (2022).

35. Kibble, T. W. Topology of cosmic domains and strings. *J. Phys. A-Math. Gen.* **9**, 1387 (1976).

36. Zurek, W. H. Cosmological experiments in superfluid helium? *Nature* **317**, 505-508 (1985).

37. Bowick, M. J., Chandar, L., Schiff, E. A. & Srivastava, A.M. The cosmological Kibble mechanism in the laboratory: string formation in liquid crystals. *Science* ***263***, 943-945 (1994).

38. Smalyukh, I. I. Knots and other new topological effects in liquid crystals and colloids. *Rep. Prog. Phys.* **83**, 106601 (2020).

39. Smalyukh, I. I. Liquid Crystal Colloids. *Annu. Rev. Condens. Matter Phys.* **9**, 207-226 (2018).

40. Park, S., Liu, Q. and Smalyukh. I. I. Colloidal surfaces with boundary, apex boojums and nested self-assembly in pyramidal-cone nematic colloids. *Phys. Rev. Lett.* **117**, 277801 (2016).

41. Müller, D., Kampmann, T. A., Kierfeld, J. Chaining of hard disks in nematic needles: particle-based simulation of colloidal interactions in liquid crystals. *Scientific Reports* **10**, 12718 (2020).

42. Jamali, V., Behabtu, N., Senyuk, B., Lee, J. A., Smalyukh, I. I., van der Schoot, P., and Pasquali, M. Experimental realization of crossover in shape and director field of nematic tactoids. Phys. Rev. E **91**, 042507 (2015).

43. Gârlea, I. C., Dammone, O., Alvarado, J., Notenboom, V., Jia, Y., Koenderink, G. H., Aarts, D. G. A. L., Lettinga, M. P., Mulder, B. M. Colloidal Liquid Crystals Confined to Synthetic Tactoids. *Scientific Reports* **9**, 20391 (2019).

44. Keber, F. C., Loiseau, E., Sanchez, T., DeCamp, S.J., Giomi, L., Bowick, M.J., Marchetti, M. C., Dogic, Z., Bausch, A. R. Topology and dynamics of active nematic vesicles. *Science* **345**, 1135-1139.

45. Liu, S., Shankar, S., Marchetti, M.C., Wu, Y. Viscoelastic control of spatiotemporal order in bacterial active matter. *Nature* **590**, 80-84 (2021).

46. Kosa, T., Sukhomlinova, L., Su, L., Taheri, B., White, T. J., Bunning, T. J. Light-induced liquid crystallinity. *Nature* **485**, 347–349 (2012).

47. Zhang, R., Redford, S. A., Ruijgrok, P.V., Kumar, N., Mozaffari, A., Zemsky, S., Dinner, A. R., Vitelli, V., Bryant, Z., Gardel, M. L., de Pablo, J. J. Spatiotemporal control of liquid crystal structure and dynamics through activity patterning. *Nature Materials* **20**, 875–882 (2021).

48. Lemma, L. M., Varghesea, M., Ross, T.D., Thomsond, M., Baskaran, A., Dogic, Z. Spatio-temporal patterning of extensile active stresses in microtubule-based active fluids. *PNAS Nexus* **2**, 1–10 (2023).

49. Faluweki, M. K., Cammann, J., Mazza, M. G., Goehring, L. Active Spaghetti: Collective Organization in Cyanobacteria. Preprint arXiv:2301.11667 (2023).

50. Zhang, R., Mozaffari, A., de Pablo, J. J. Towards Autonomous Materials from Active Liquid Crystals. *Nature Reviews Materials* **6**, 437-453 (2021).

51. Yokoo, R., Hood, R.D., Savage, D.F. Live-cell imaging of cyanobacteria. *Photosynth. Res.* **126**, 33-46 (2015).

52. Born, M. &Wolf, E. Principles of Optics: Electromagnetic Theory of Propagation, Interference and Diffraction of Light (Elsevier, 2013).

53. Cordelières F. P. Manual Tracking. https://imagej.nih.gov/ij/plugins/track/Manual%20Tracking%20plugin.pdf

54. Nussbaum-Krammer, C.I. et al. Investigating the spreading and toxicity of prion-like proteins using the metazoan model organism C. elegans. *J. Vis. Exp.* **95**, p.e52321 (2015).

55. Ebbens, S. et al. Self-assembled autonomous runners and tumblers. *Phys. Rev. E* **82**, 015304 (2010).

**Acknowledgments**

We thank E. B. Johnson, T. Lee, B. Senyuk, J.-S. Wu for discussions, and Elizabeth Trower for sharing the *Geitlerinema* sp. culture. We acknowledge the support of the US Department of Energy grants DE-SC0019306 (A.R., C.G., J.C.C., I.I.S.) and DE-SC0020361 (A.R., J.C.C., I.I.S.). I.I.S. acknowledges hospitality of the International Institute for Sustainability with Knotted Chiral Meta Matter (SKCM$^2$) at Hiroshima University and Kavli Institute for Theoretical Physics at the University of California at Santa Barbara during his sabbatical visits, where he was partly discussing research findings and working on this article.

**Author contributions**

I.I.S. and J.C.C. conceived and designed the project. A.R. and C.G. performed the bacteria growth. I.I.S. and A.R. designed the experiments. A.R and C.G. performed the microscopy imaging. A.R. analyzed the data. I.I.S., A.R., C.G., and J.C.C. interpreted the results. I.I.S. and A.R. wrote the manuscript.

**Competing interests**

J.C.C. is a co-founder and holds equity in Prometheus Materials. All other authors declare that they have no competing interests.

**Additional information**

**Supplementary information** The online version contains supplementary material available at

**Correspondence** and requests for materials should be addressed to Ivan I. Smalyukh.

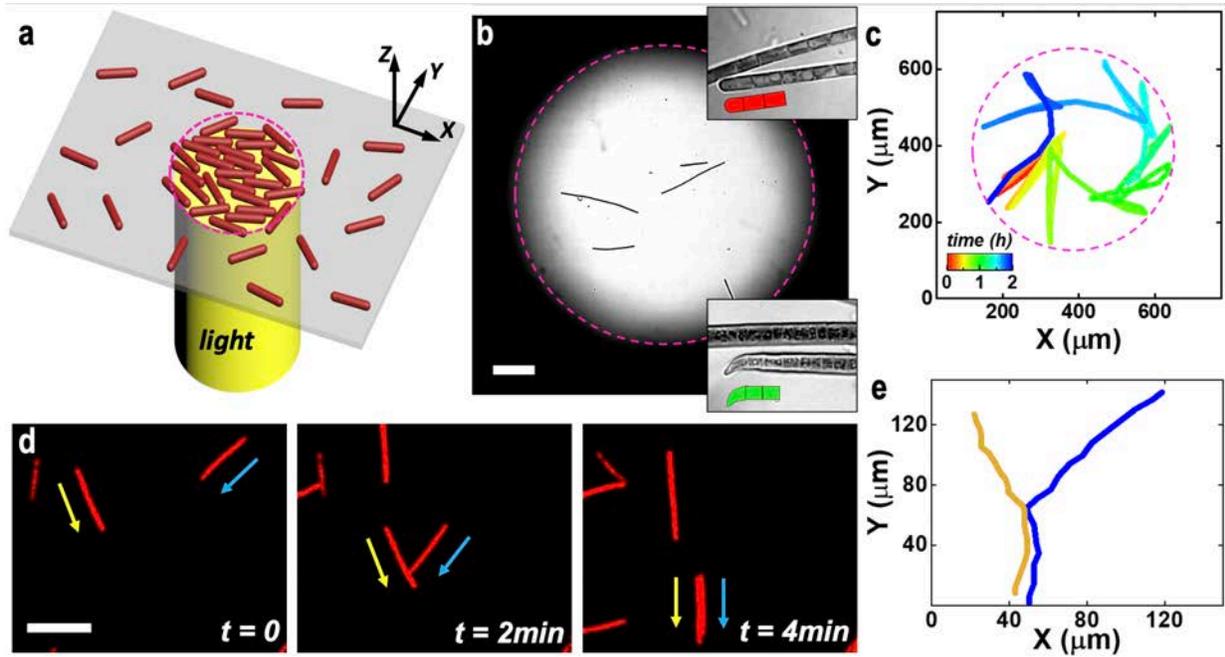

**Fig. 1 | 2D confinement, phototactic behavior and interactions of trichomes. a,** Schematic of bacterial response to light with intensity gradients. **b,** Optical micrograph of trichomes in a locally illuminated circular region; insets show higher magnification, where *Geitlerinema* sp. (top) and *Oscillatoria brevis* (bottom) filaments are seen while being optically resolved at a single cell level; the red and green models depict the chain-like associations of cyanobacteria at the end of the filaments of two different species, with the bacteria at the ends taking geometric forms different from the ones in the middle of filaments. **c,** Motility confined within the illuminated area corresponding to **b**. **d,** Cyanobacterial filaments and their collision followed by co-alignment, as revealed by optical imaging; elapsed time *t* is marked in the bottom-left corners. **e,** Corresponding motion traces depicting propulsion directionalities before and after the collision. White scale bars, 40 μm.

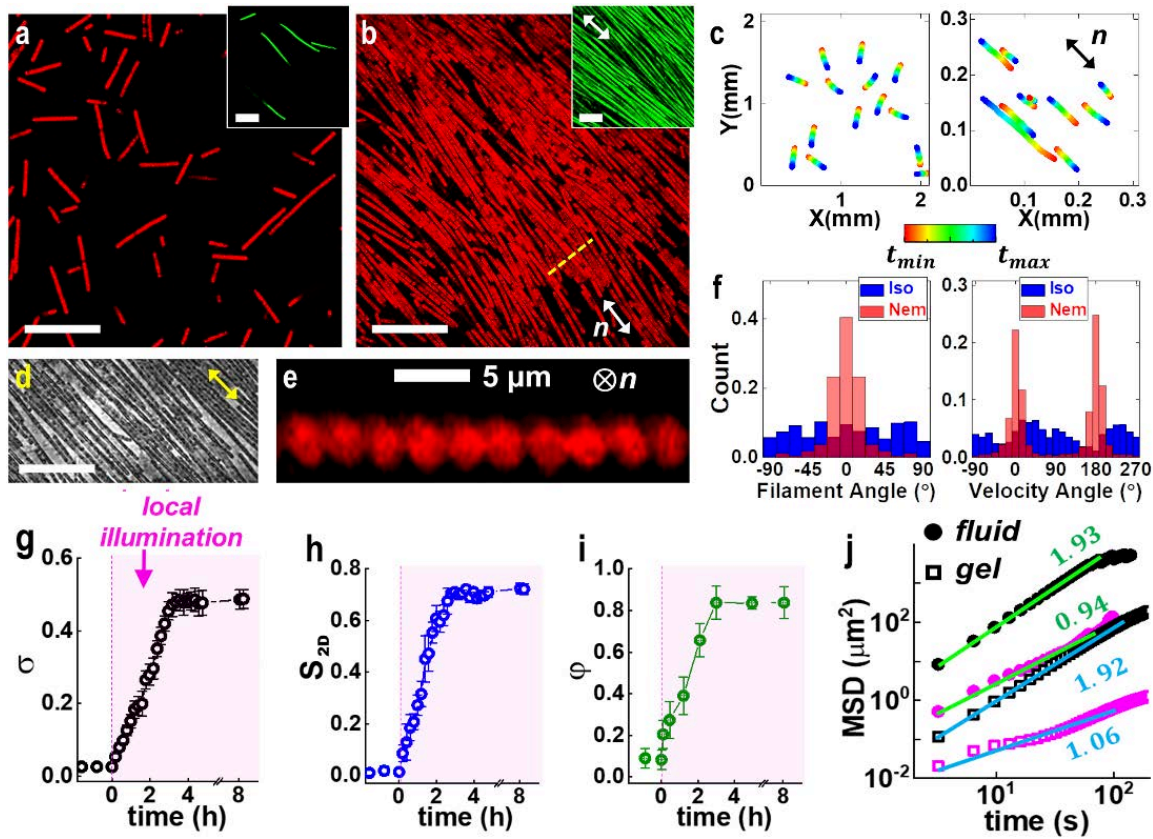

**Fig. 2 | Emergence of 2D phototactic active nematics. a-b,** Disordered and nematic morphologies of *Geitlerinema* sp. and *O. brevis* (insets) before (**a**) and after (**b**) illumination. **c,** 2D particle trajectories in disordered (left) and nematic (right) states, in the latter case revealing bipolar nematic-like motions of filaments. Time elapsed since the beginning of tracking ($t_{min}$) until the end of tracking ($t_{max}$) is depicted using the color scale shown in the bottom inset, where $t_{max}-t_{min}$=5min. **d,** Nematic of motile trichomes entrapped at the water-air interface, showing emergence of the 2D active nematic for a different 2D confinement method. **e,** Depth-resolved confocal-like nonlinear optical imaging cross-section of 2D one-trichome—thick nematic bacterial community acquired from a region in (**b**) along the yellow dashed line. **f,** Distributions of filament orientations (left) and of their velocity vectors (right) in the disordered and nematic states, with the latter emerging from the phototactic cyanobacterial response. **g-i,** Time evolution of trichomes relative surface area coverage $\sigma$, orientational $S_{2D}$ and velocity $\varphi$ order parameters during continuous spatially localized illumination. **j,** MSDs parallel (black) and perpendicular (pink) to **n** for the fluid and gel states, probed over short time scales ~100s. Scale bars, 40 μm, if not mentioned otherwise. Error bars are the standard deviations (see Methods).

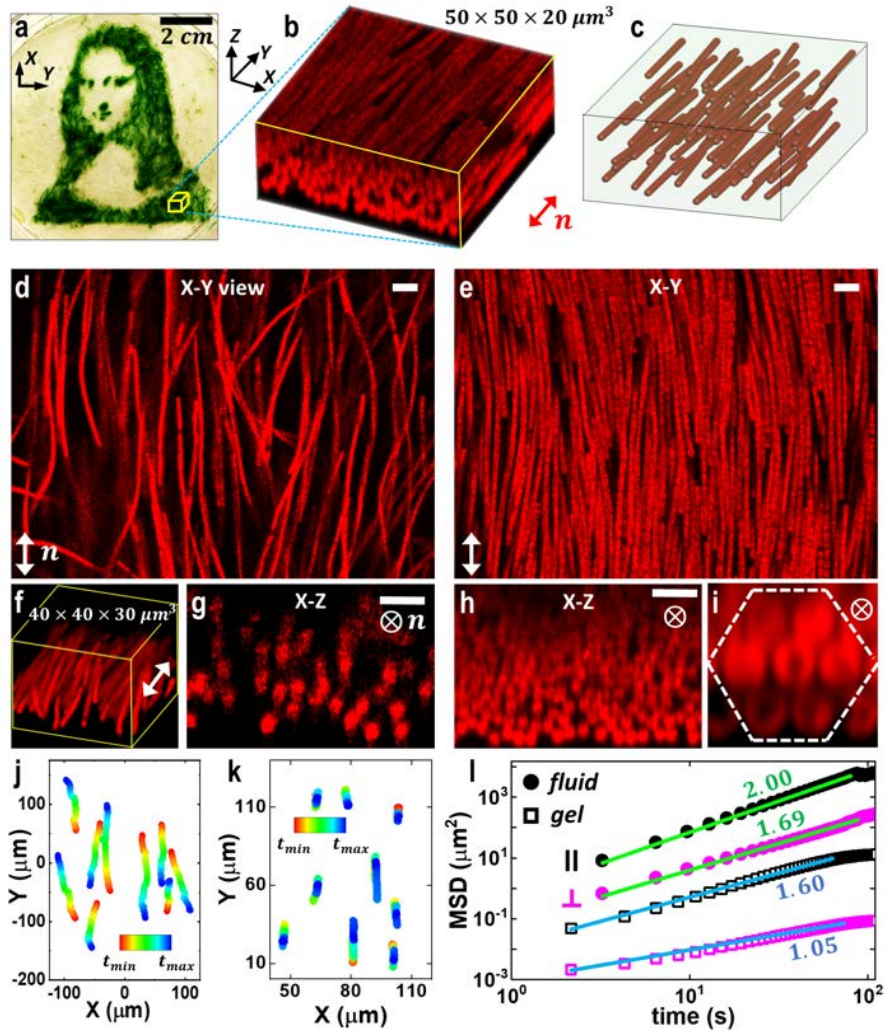

**Fig. 3 | 3D phototactic active nematics. a,** Image of a 3D *Geitlerinema* sp. community induced by projecting a pattern of light with intensity gradients, obtained by projecting an inverse intensity pattern corresponding to the Mona Lisa painting of Leonardo da Vinci. **b,** Zoomed in 3D nematic gel ordering of trichomes captured by multiphoton imaging. **c,** 3D schematic of nematically arranged filaments embedded in the polysaccharide slime. **d-e,** Cross-section of the 3D nematic fluid (**d**) and gel (**e**) parallel to **n**. **f,** a 3D nonlinear multiphoton-absorption-based optical image showing the nematic fluid ordering. **g,** Multiphoton-absorption-based cross-sectional image of a nematic fluid taken perpendicular to **n**. **h-i,** Cross-sectional images of nematic gel obtained for a vertical plane perpendicular to **n**. The images reveal nematic liquid-crystalline order without positional correlations of bacterial filaments (**h**) and locally hexagonal near-neighbour arrangement within such large communities, which is sometimes seen near the surface closest to the illumination source (**i**). **j-k,** Typical filament trajectories in the 3D nematic fluid (**j**) and gel (**k**) media, demonstrating bipolar particle propulsion parallel to **n**. Time elapsed since the beginning of tracking ($t_{min}$) until the end of tracking ($t_{max}$) is depicted using the color scale shown in the insets, where, where $t_{max}$-$t_{min}$=10min for (**j**) and 30min for (**k**). **l,** MSDs parallel (black) and perpendicular (pink) to **n** for the fluid and gel states. Scale bars, 10μm, if not mentioned otherwise.

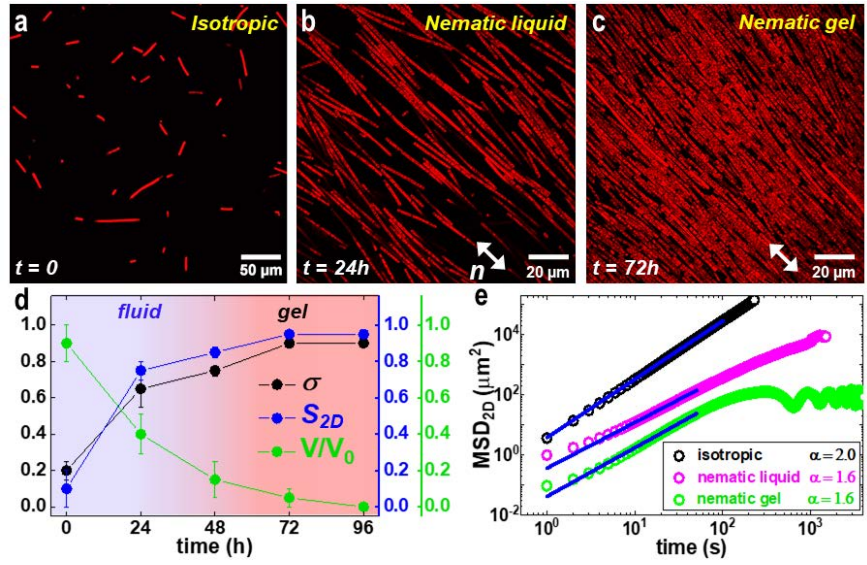

**Fig. 4 | Fluid-to-gel transition in cyanobacterial communities with the nematic order. a-c**, Time evolution of 2D *Geitlerinema* sp. assembly during continuous local illumination over 72 hours. **d,** Temporal evolution of filaments' relative surface area coverage $\sigma$, orientational order parameter $S_{2D}$, and normalized velocity $V/V_0$, where $V_0 = 2$ μm/s. Over time, in the course of days, the polysaccharide density continuously builds up, prompting the contiguous transition from the nematic fluid state (bluish background) to the gel state (pink background) state. **e,** Typical MSD curves corresponding to filament motility for an as-prepared sample starting at elapsed time *t = 0* (black), at *t = 24 h* (pink) and *t = 72 h* (green), where the oldest sample shows gel-like behavior with quasi-periodic motility.

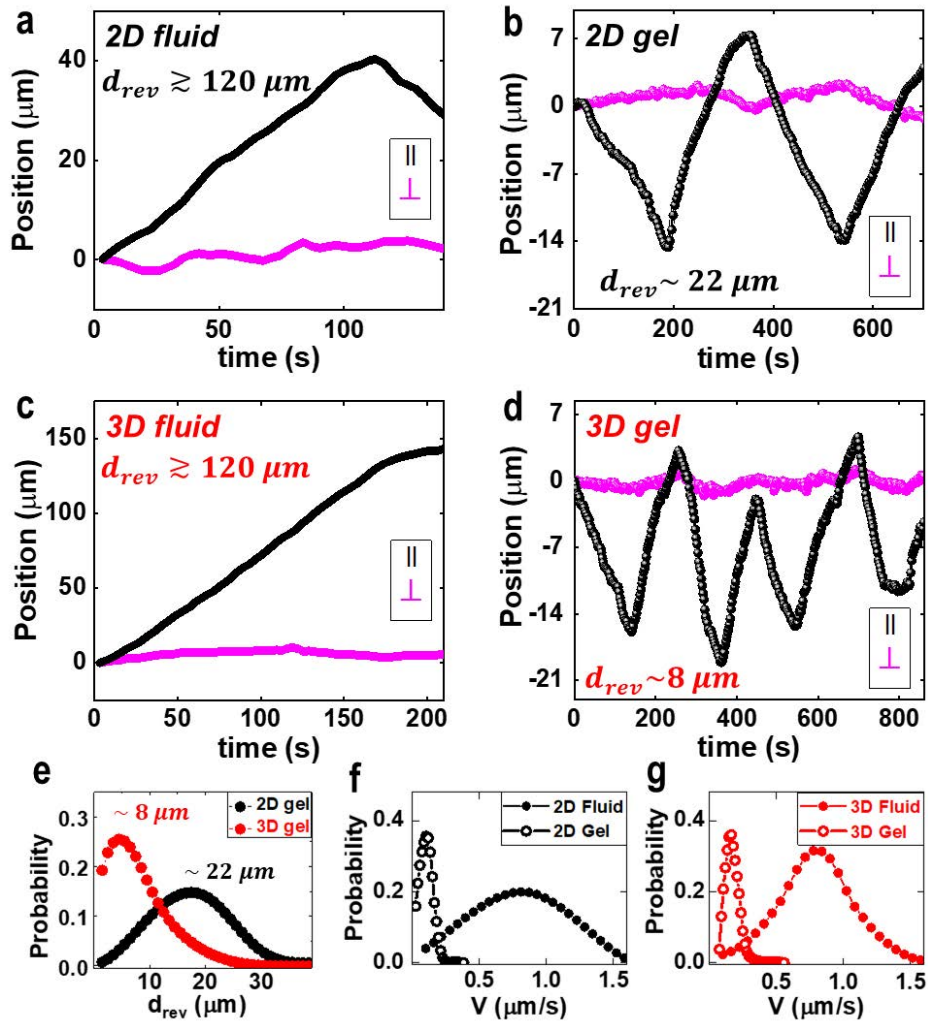

**Fig. 5 | Trajectories, motion reversals and velocity distributions in nematic fluids and gels.**
**a-b,** Parallel (black) and perpendicular (pink) position versus time trajectories of *Geitlerinema* sp. cyanobacteria for the 2D fluid and gel states, respectively. **c-d,** Similar data for the 3D nematic fluid and gel, respectively. Average distances between every two consecutive reversal points along the same filament trajectories $d_{rev}$ are estimated and marked on different plots. **e,** Probability plots for $d_{rev}$ in 2D and 3D nematic gels formed by *Geitlerinema* sp. bacteria, with the mean values of $d_{rev}$ found to be 22 μm and 8 μm, respectively. **f-g,** *Geitlerinema* sp. filaments' velocities plots for the (**f**) 2D fluid and gel, and (**g**) for the 3D fluid and gel. The data reveal faster filament propulsion in the fluid state (solid circles) as compared to the gel state (open circles), with velocities of 0.8 μm/s versus 0.1 μm/s in the 2D case, and 0.84 μm/s versus 0.13 μm/s in the 3D case.

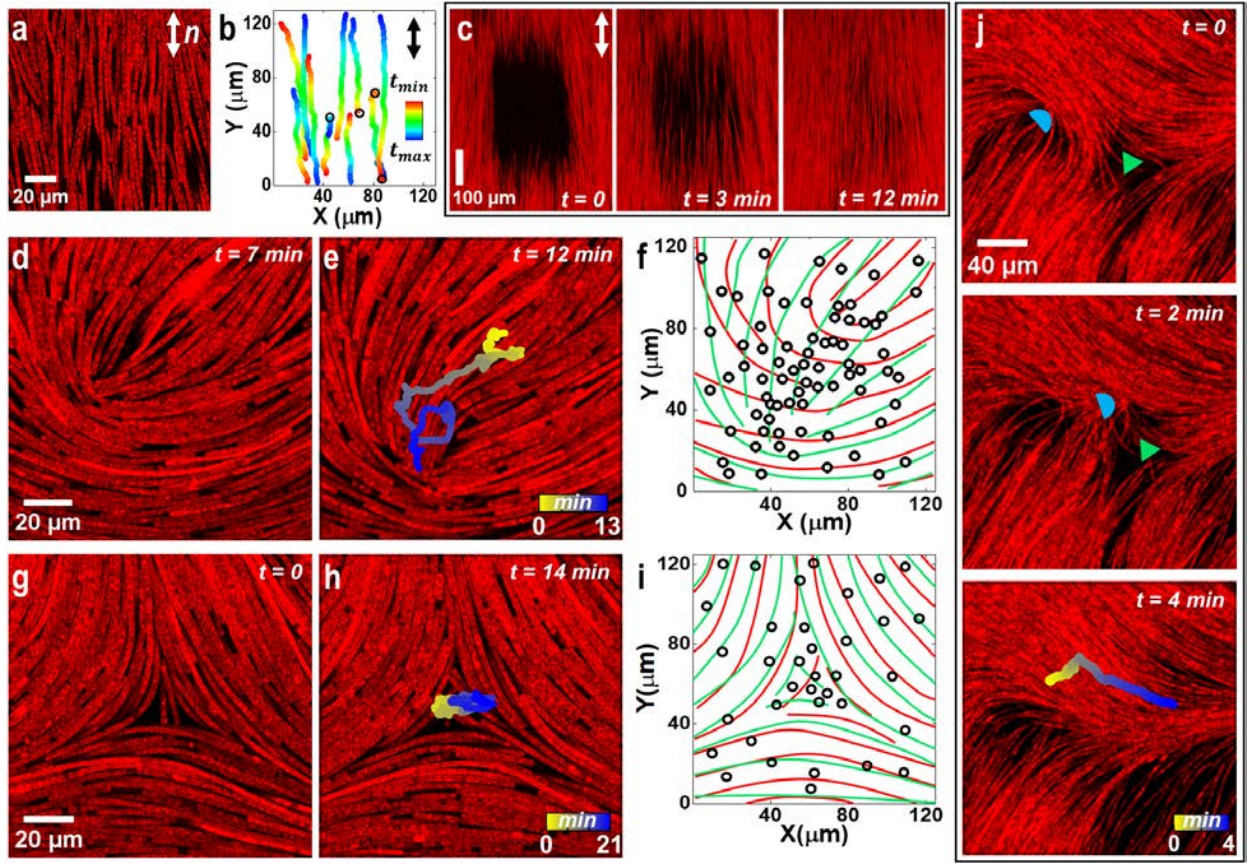

**Fig. 6 | Fluid state of 2D phototactic nematics without and with defects. a,** Image of a 2D nematic monodomain fluid assembled from *Geitlerinema* sp. visualized at the single cell level, where conjoined individual cells are visible. **b,** Color-coded spatiotemporal filaments' trajectories in the nematic fluid, where black open circles are the propulsion reversal points. Time elapsed since the beginning of tracking ($t_{min}$) until the end of tracking ($t_{max}$) is depicted using the color scale shown in the inset, where $t_{max}$-$t_{min}$=10min. **c,** Temporal evolution of a photobleached monodomain region within a nematic fluid. **d-e,** Temporal evolution of the +1/2 defect displacement in a nematic fluid indicated by its color-coded trajectoryI (**e**), with the elapse time given in minutes and the frames in **d** and **e** shown for elapsed times marked on frames. **f,** Reversal points and director field evolution over 20 minutes illustrated by red (original) and green (evolved) lines for the moving +1/2 defect. **g-h,** Virtually negligible fluctuation-like displacement of −1/2 defect in fluid shown by its pointlike time-color-coded trajectory oI (**h**). **i,** Reversal points and small director field evolution for the −1/2 defect. **j,** Annihilation of the moving +1/2 (blue semicircle) and static −1/2 (green triangle) defects, leaving behind a defect-free deformed nematic state after the elapsed time of $t \approx 4\ min$. Colored trajectory at $t = 4\ min$ is the +1/2 defect's spatiotemporal trace toward annihilation.

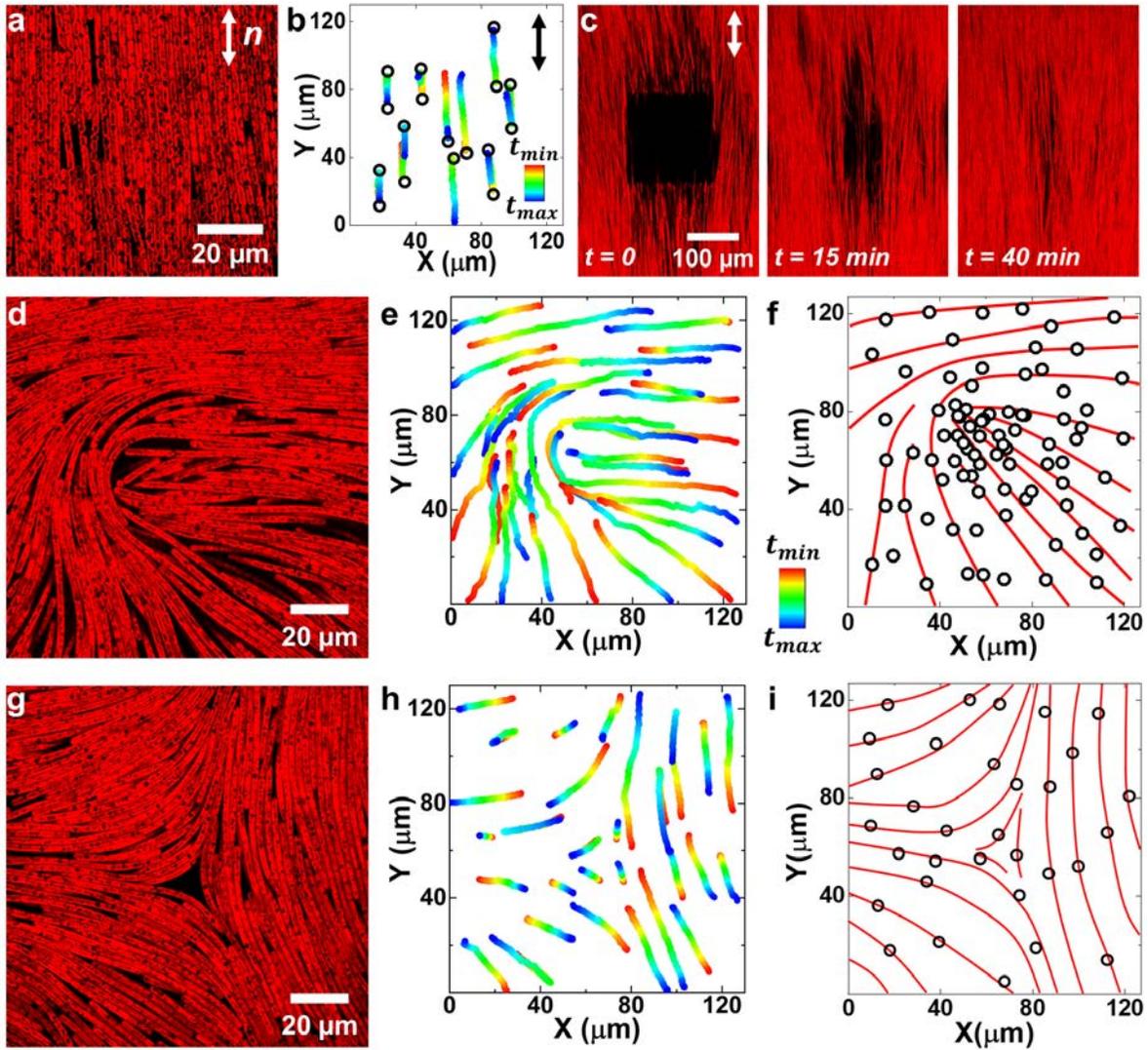

**Fig. 7 | Gel states of 2D phototactic nematics without and with defects. a,** Nonlinear optical imaging of a 2D nematic monodomain gel. **b,** Filament trajectories in the gel are confined between the propulsion reversal points (black open dots). **c,** Temporal evolution of a photobleached region in a 2D nematic gel. **d,** an immobile +1/2 defect in the gel state. **e,** color-coded filament trajectories around the static "frozen" +1/2 defect core. **f,** Static director field in the gel state near the +1/2 defect overlaid with the reversal points. **g-i,** Immobile −1/2 defect, filament trajectories, and the static director field around it, respectively. Time elapsed since the beginning of tracking ($t_{min}$) until the end of tracking ($t_{max}$) is depicted using the color scale shown in the insets of (**b, e** and **h**), where $t_{max}$- $t_{min}$=20min in (**b**) and 10min in (**e,h**).

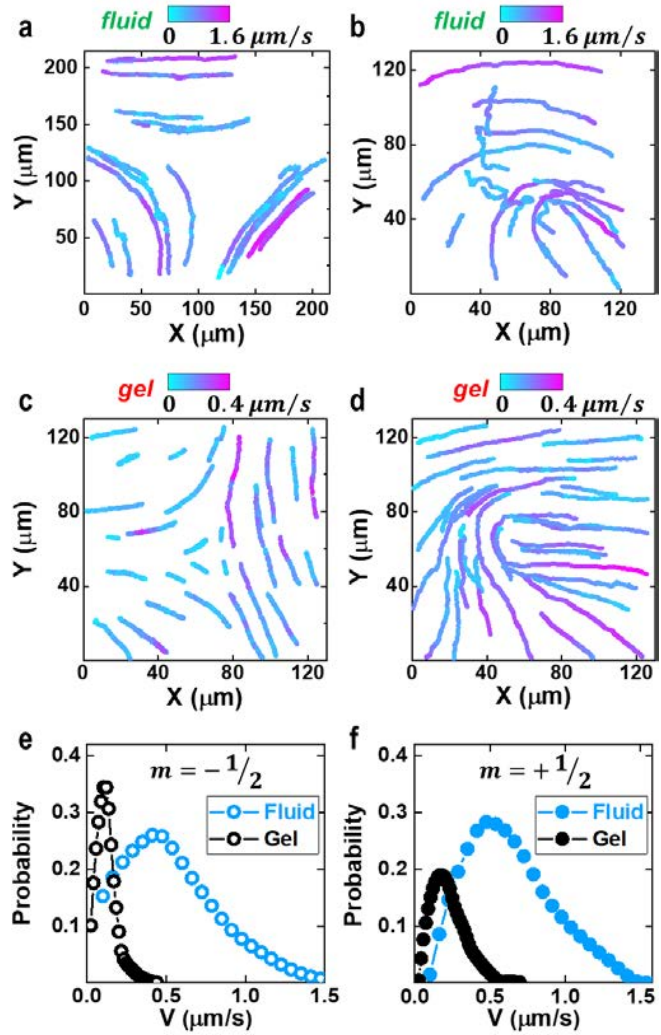

**Fig. 8 | Bacterial dynamics and velocity probability distributions near topological defects. a-d,** Typical filament trajectories around (**a**) −1/2 and (**b**) +1/2 defects in the nematic fluid and gel states (**c** and **d**, respectively). Color coding shows the local variations of filament velocity according to the color schemes shown in the insets. In the fluid state, the filaments propel faster and along trajectories continuously evolving with time, differently from their gel counterparts. **e-f,** *Geitlerinema* sp. velocity histograms extracted from (**a-d**), showing filament velocities around (**e**) −1/2 and (**f**) +1/2 defects in fluid and gel states, revealing faster filament propulsion in the fluid state (blue) as compared to the gel state (black), with corresponding velocities within 0.5-0.6 µm/s for nematic fluids versus 0.1-0.2 µm/s for nematic gels.

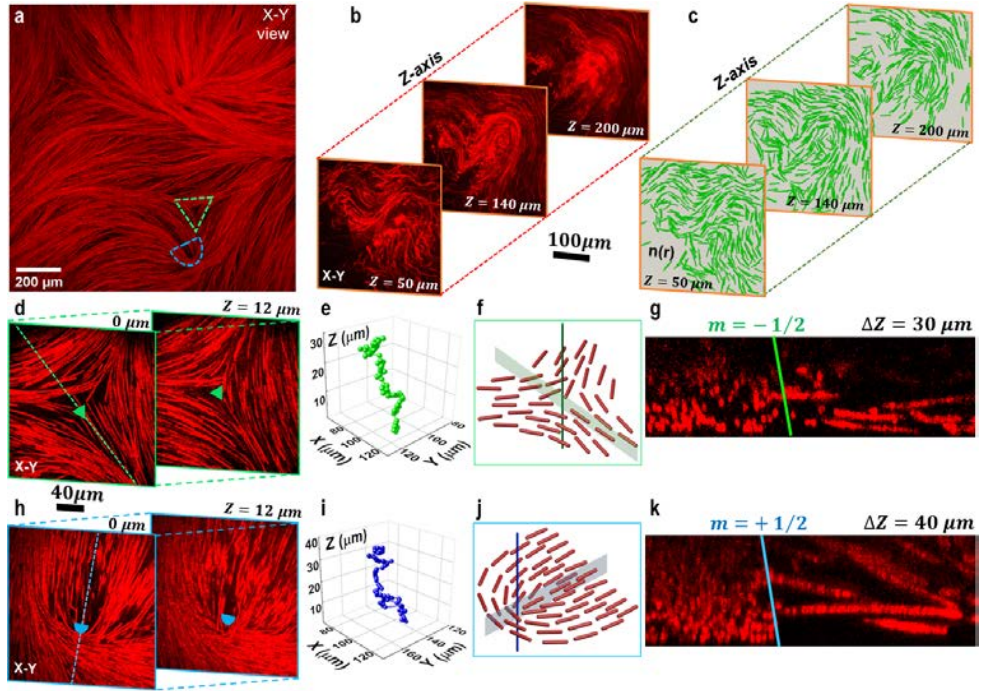

**Fig. 9 | Polydomain 3D phototactic nematic liquid crystal. a,** Large-field nonlinear optical image of the *Geitlerinema* sp. polydomain 3D nematic sample with disclinations having local 2D cross-sectional structures characterized by the local configurations corresponding to −1/2 (green triangle) and +1/2 (blue semicircle) defects. **b,** Cross-sectional nonlinear optical images obtained at different sample depths Z in a thick polydomain nematic fluid. **c,** reconstructed 2D projections of **n(r)** in different cross-sectional planes, revealing a vortex line that is locally structurally similar to the +1/2 defect at sample depths Z = 140 μm and Z=200 μm. **d,** Depth-resolved nonlinear optical images at different depths revealing the 3D structure of the defect line meandering through the active nematic's bulk. **e,** The corresponding geometric configuration of the vortex line's core (green) revealed by nonlinear optical scanning images at different sample's depths Z. **f,** Schematic showing the filament orientation structure and **n(r)** (green) around the vortex line with the local −1/2 defect structure. The green shaded cross-sectional plane corresponds to the green dashed cross-sectional line in (**d**) passing through the defect core. **g,** Multiphoton-absorption-based nonlinear optical cross-sectional image demonstrating the filament orientation in the green shaded plane schematically shown on the image in (**f**). The green solid line in (**g**) indicates the position and orientation of the disclination line within the nematic bulk. **H,** The 3D structure of a line defect with local configuration similar to the +1/2 2D defect. **i,** the 3D geometric path/morphology of this line defect. **j,k,** Schematic (**j**) and (**k**) cross-sectional (**k**) image of +1/2 defect, where the blue shaded plane corresponds to the cross-sectional plane passing through the defect core, and solid blue line is disclination line. **j,** Schematic showing the filament orientation structure and **n(r)** (green) around the vortex line with the local +1/2 defect structure. The blue shaded cross-sectional plane corresponds to the green dashed cross-sectional line in (**h**), passing through the defect core. **k,** Multiphoton-absorption-based nonlinear optical cross-sectional image demonstrating the filament orientation in the blue shaded plane schematically shown on the image in (**h**).

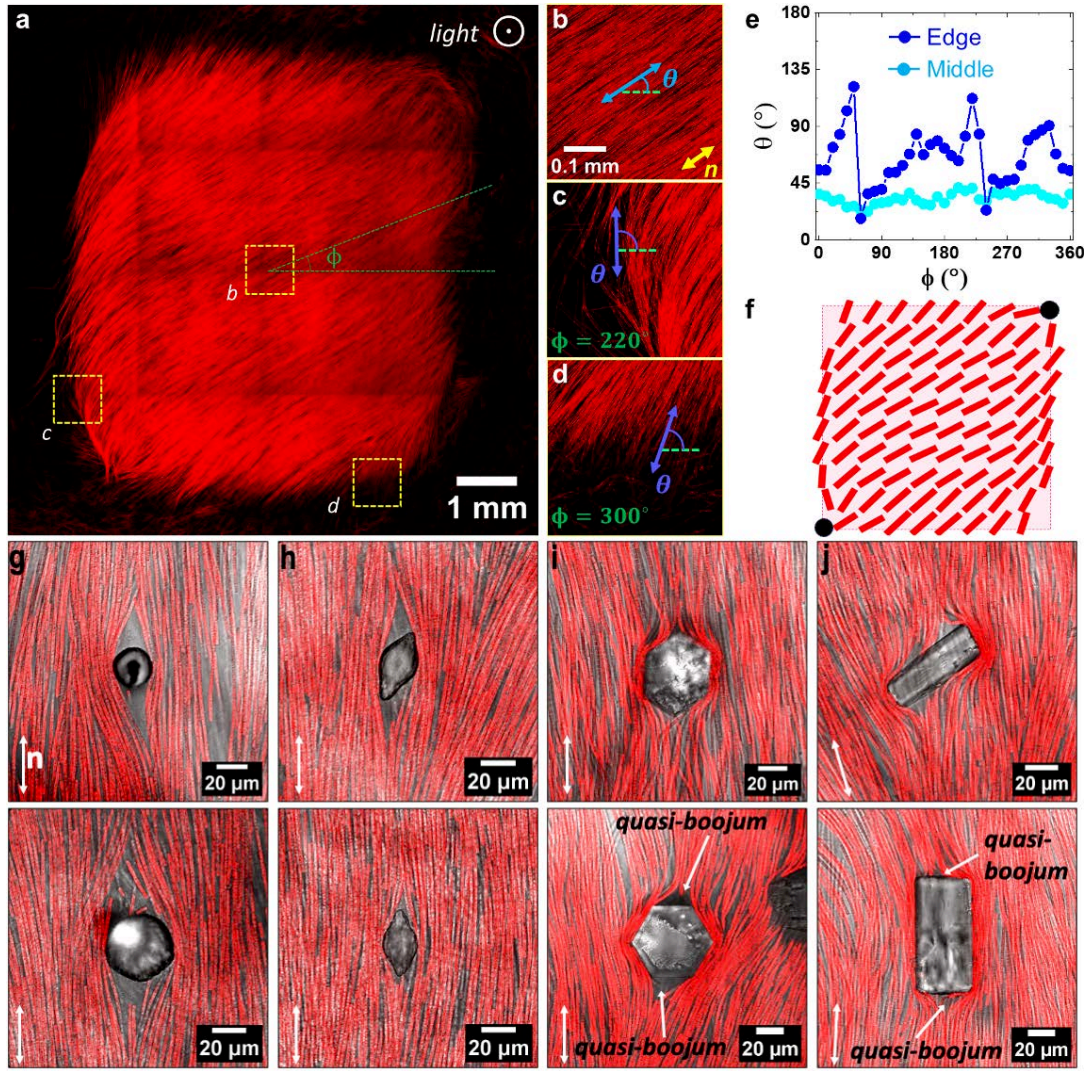

**Fig. 10 | Phototactic tactoid and foreign inclusions in the cyanobacterial active nematic. a,** Nonlinear optical image of a phototactic tactoid that emerged above the glass surface due to continuous square-shaped illumination with dimensions 5x5 mm$^2$. **b,** Zoomed-in image of the uniform nematic alignment in the middle of the tactoid-like domain. Angle $\theta$ defines the director orientation. **c-d,** Zoomed-in images showing the director orientation at the edge of the domain at the different circumnavigation angles $\varphi$. **e,** director orientation versus circumnavigation angle in the middle (cyan) along the perimeter of the box **b** and along the edge (blue) of the entire domain. **f**, schematics showing the tactoid-like bipolar director field and underlying square-shaped light source; red rods represent filament orientations and black filled circles indicate the location of quasi-boojum-like defects. **g-j,** Inclusions in the cyanobacterial nematic that emerge during the slow and controllable nutrition medium drying from *Geitlerinema* sp. bacterial communities, where salt residues crystallize into the objects with shapes of (from left to right) (**g**) disks, (**h**) spindles, (**i**) hexagons, and (**j**) rectangles. Active nematic analogs of boojums are observed, and are better pronounced for larger inclusions, as indicated with arrows in the bottom images of (**j**) and (**k**). As expected, due to genus-zero (Euler characteristics of 2) confining surfaces, two quasi-boojums are observed for both tactoid-like domains and spherical inclusions,[38-42] both imposing weak tangential boundary conditions for **n**.

# Supplementary information

## Photosynthetically-powered phototactic active nematic fluids and gels

**Authors:**

Andrii Repula[1], Colin Gates[2,3], Jeffrey C. Cameron[2,3] and Ivan I. Smalyukh[1,2,4,5*]

**Affiliations:**

[1]Department of Physics, University of Colorado, Boulder, Colorado 80309, USA

[2]Renewable and Sustainable Energy Institute, National Renewable Energy Laboratory and University of Colorado, Boulder, Colorado 80309, USA

[3]Department of Biochemistry, University of Colorado, Boulder, Colorado 80309, USA

[4]International Institute for Sustainability with Knotted Chiral Meta Matter (WPI-SKCM²), Hiroshima University, Higashihiroshima, Hiroshima, 739-8526, Japan

[5]Materials Science and Engineering Program and Department of Electrical, Computer and Energy Engineering, University of Colorado, Boulder, Colorado 80309, USA

*Corresponding author. Email: ivan.smalyukh@colorado.edu

**Formation of phototactic active nematics by different bacterial species**

We conducted our experiments on phototactic active nematics utilizing two types of cyanobacteria: *Geitlerinema* sp. and *O. brevis*. Taxonomically, these bacteria belong to the same order, Oscillatoriales. However, within the Oscillatoriales they are rather divergent and beyond being filamentous they are morphologically disparate. From the physical point of view, their cells are considerably different in size and shape: the *Geitlerinema* sp. cell has the shape of a blunt rod, similar to a unicellular bacillus, whereas *O. brevis* cells are disklike and most of their surface area is in contact with other cells in their filament (Supplementary Fig.1). Based on our microscopy observations, the *Geitlerinema* sp. filament effective persistent length (in reference to activity-driven bending tendency of filaments, differently from the conventional persistence length concept used in polymer science) is estimated to be 70 μm, whereas that of *O. brevis* is around 40 μm. Effectively, a filament of *Geitlerinema* sp. contains fewer cells on average and thus has fewer junctions between cells for flexibility but also has more freedom to flex due to the aspect ratio of the cells, whereas *O. brevis* has many joints but can barely exploit their rotational degrees of freedom individually due to the cell diameter exceeding its length. Moreover, we observed different shapes of the filament terminal cell (Supplementary Fig. 1b,c): *Geitlerinema* sp. trichomes have a rounded terminal shape, whereas *O. brevis* exhibits a pointed tip. Filament velocity measurements in highly diluted states on a glass surface indicate that *Geitlerinema* sp. is roughly twice as fast as *O. brevis* (Supplementary Fig.1h). Despite these differences between *Geitlerinema* sp. and *O. brevis*, both cyanobacterial species can form phototactic active nematic states under appropriate illumination conditions. From the biological standpoint, along with the phototaxis, these emergent ordered phototactic active nematic states provide optimal photosynthetic conditions for filaments under dense packing conditions, enabling efficient cell metabolism across the colony. Furthermore, nematic morphologies based on the two active light-powered particle types reveal topological defects (Supplementary Figs. 6 and 7). Thus, the cyanobacterial trichomes' assembly into active nematic states under localized illumination may be a common property of many types of filamentous cyanobacteria, albeit further studies will be needed to verify this and to explore how things like shape chirality may potentially influence collective behavior.

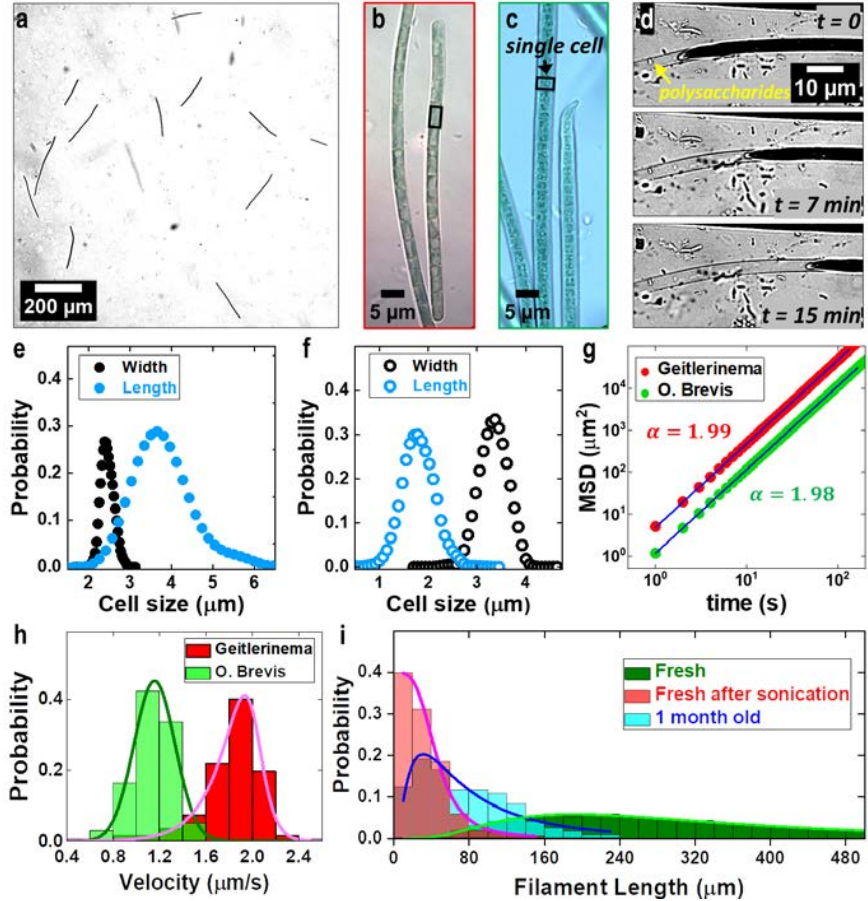

**Supplementary Fig. 1 | Cyanobacteria cell and filament dimensions and motility.**

**a,** Large-scale microphotograph of freshly grown cyanobacteria filaments. **b-c,** Optical images of *Geitlerinema* sp. (**b**) and *Oscillatoria brevis* (*O. brevis*) (**c**) cyanobacteria live colonially in long filaments (trichomes) consisting of multiple conjoined cells with anisotropic shapes. **d,** Successive high magnification brightfield optical images showing a self-propelled cyanobacterial filament as it produces and leaves behind a polysaccharide track[33]. **e-f,** Distributions for *Geitlerinema* sp. (**e**) and *O. brevis* (**f**) single cell widths with average values $a_{Geit}$ = 2.5 μm and $a_{O.br}$ = 3.4 μm, respectively. Distributions for *Geitlerinema* sp. and *O. brevis* single cell lengths (denoted *b*) with average values of $b_{Geit}$ = 3.7 μm and $b_{O.br}$ = 1.8 μm, respectively. **g,** MSDs of freshly grown *Geitlerinema* sp. (red) and *O. brevis* (green) filaments indicating their ballistic motility regime with the diffusion exponent α ~ 2 obtained via fitting[6] $MSD \propto t^{\alpha}$. **h,** Velocity distributions for the freshly grown *Geitlerinema* sp. (red) and *O. brevis* (green) filaments showing their mean values of 2 μm/s and 1.2 μm/s, respectively. Histograms corresponding to *Geitlerinema* sp. and to *O. brevis* velocities have been fit with the Asymmetric double Sigmoidal and Gaussian functions, with the fitting curves used for eye guiding. **i,** Polydisperse distributions of the entire filament length of *Geitlerinema* sp. cyanobacteria characterized for different conditions: freshly grown (green), freshly grown after 1 minute of sonication (red), and aged over 1 month following sonication (cyan). Mean filament length values of 250 μm, 40 μm, and 75 μm are found for each case respectively. Histograms describing the fresh, aged and sonicated filaments' lengths have been fit with functions for eye guiding. Fitting of all histograms is done with OriginPro 2021 software.

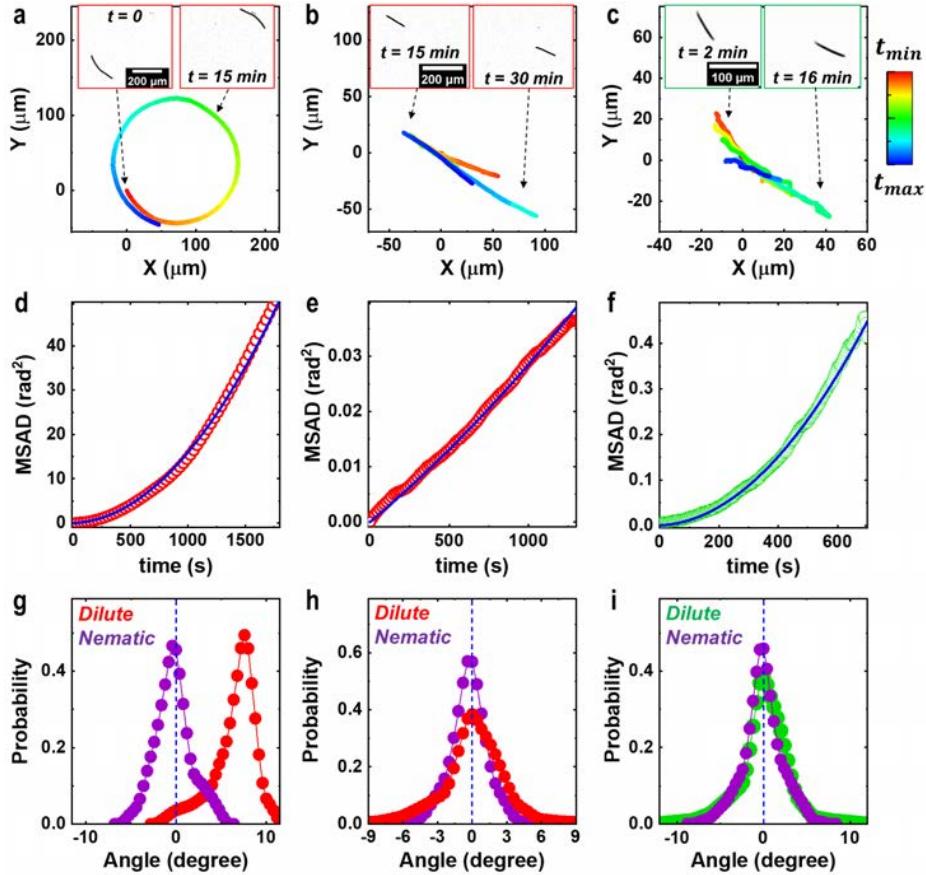

**Supplementary Fig. 2 | Filament trajectory types.**
**a-c,** Typical color-coded traces acquired over several hours of imaging of (**a**) long naturally grown >200μm and (**b**) shorter <200μm (obtained with the help of sonication) cyanobacterial filaments *Geitlerinema* sp., as well as (**c**) a similar trajectory of *O. brevis*. Time elapsed since the beginning of tracking ($t_{min}$) until the end of tracking ($t_{max}$) is depicted using the color scale shown in the inset, where $t_{max}$-$t_{min}$=60min. **d-f,** MSAD for the longer and shorter filaments of the *Geitlerinema* sp. and *O. brevis* filaments, respectively, in the highly diluted samples. Blue lines are the best fits by the power law[42] $MSAD = 2D_r t + w^2 t^2$, where $D_r$ is the rotational diffusion coefficient measured in rad$^2$/s and $w$ is the angular velocity measured in rad/s. The fitting gives values of (**d**) $5·10^{-4}$ and $3.9·10^{-3}$ for longer *Geitlerinema* sp., (**e**) $1.2·10^{-5}$ and $6.7·10^{-5}$ for shorter *Geitlerinema* sp., and (**f**) $1.6·10^{-5}$ and $9.3·10^{-4}$ for *O. brevis*, respectively. **g-i,** Corresponding distributions of the angle difference which filaments display between two consecutive frames (that are taken 15-20 s apart) in highly diluted (red for *Geitlerinema* sp. and green *O. brevis*), and nematic states (violet for both species). Longer *Geitlerinema* sp. filaments in highly diluted state (**a**) exhibit polar motions along circular trajectory and thus reveal the highly asymmetric angular distribution (**g**).

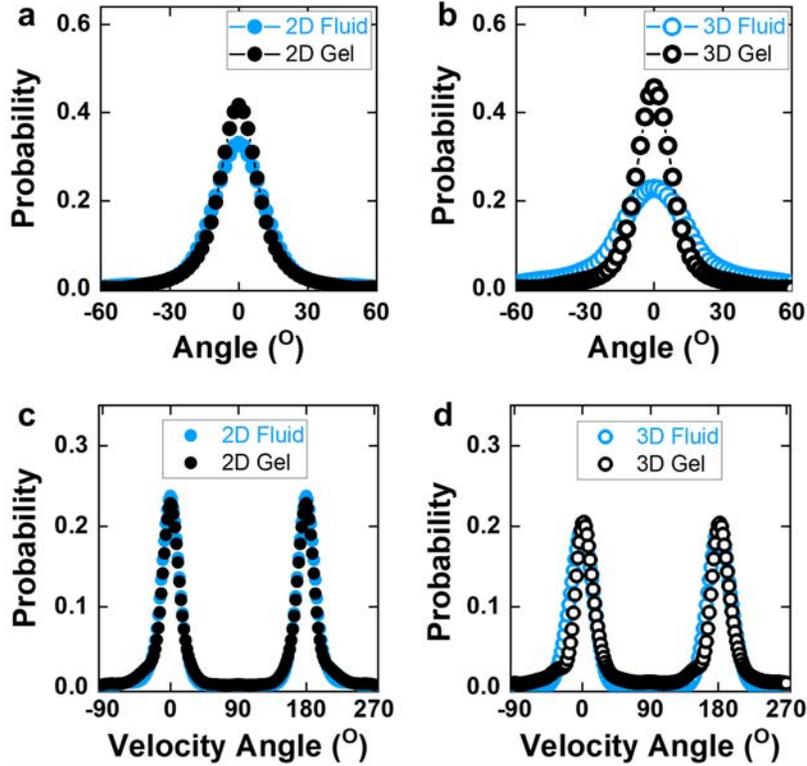

**Supplementary Fig. 3 | Distributions of filament and velocity vector orientations.**
**a-b,** *Geitlerinema* sp. angular histograms for (**a**) filaments in the 2D nematic fluid and gel states, and (**b**) for 3D fluids and gels. Orientational order parameters have been calculated based on these distributions using the numerical integration[6]. All nematic states are characterized by high orientational order parameters of typical values *0.7-0.9*. In the gel states (black), the order parameters are slightly higher compared to their fluid (blue) counterparts, consistent with the denser filament packing. **c-d,** *Geitlerinema* sp. filament velocity angular histograms for (**c**) 2D fluid and gel, and (**d**) for 3D fluid and gel. Velocity order parameters $\varphi$ have been calculated from these distributions using the expression for 2D nematic system[7]. All nematic states reveal high velocity order parameter values within the range of 0.7-0.9.

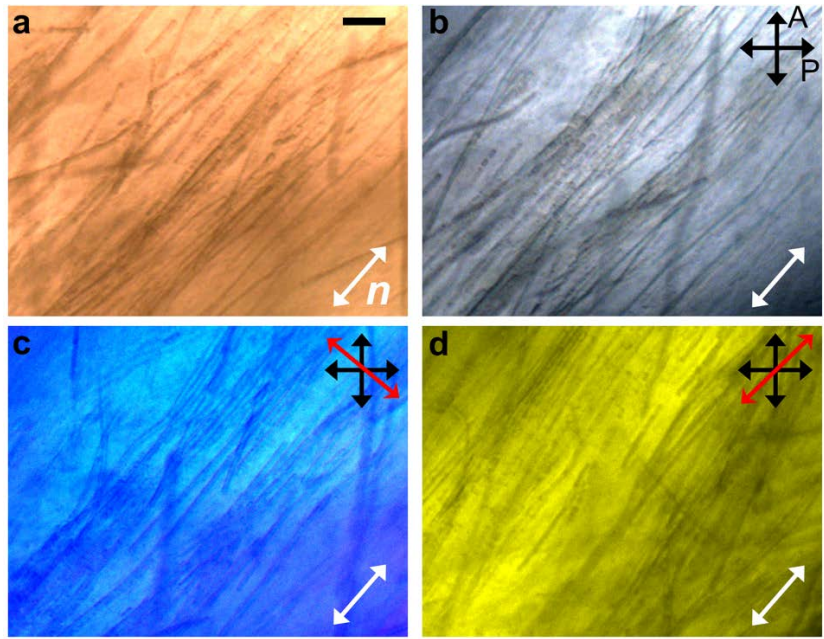

**Supplementary Fig. 4 | Optical micrograps of 3D phototactic nematic samples.** The images reveal presence of polysaccharide slime nematic ordering, in addition to orientationally ordered filaments. **a-d,** Optical microscopy images in the brightfield (**a**) and polarizing (**b**) modes for 3D nematic samples with cyanobacterial filaments embedded in slime within the nematic state, where some images are obtained with the additional retardation plate inserted with its slow axis perpendicular (**c**) and parallel (**d**) to **n**. Black double arrows illustrate the orientation of polarizers while the red ones are the retardation plate's slow axis orientations. Scale bar, 10μm.

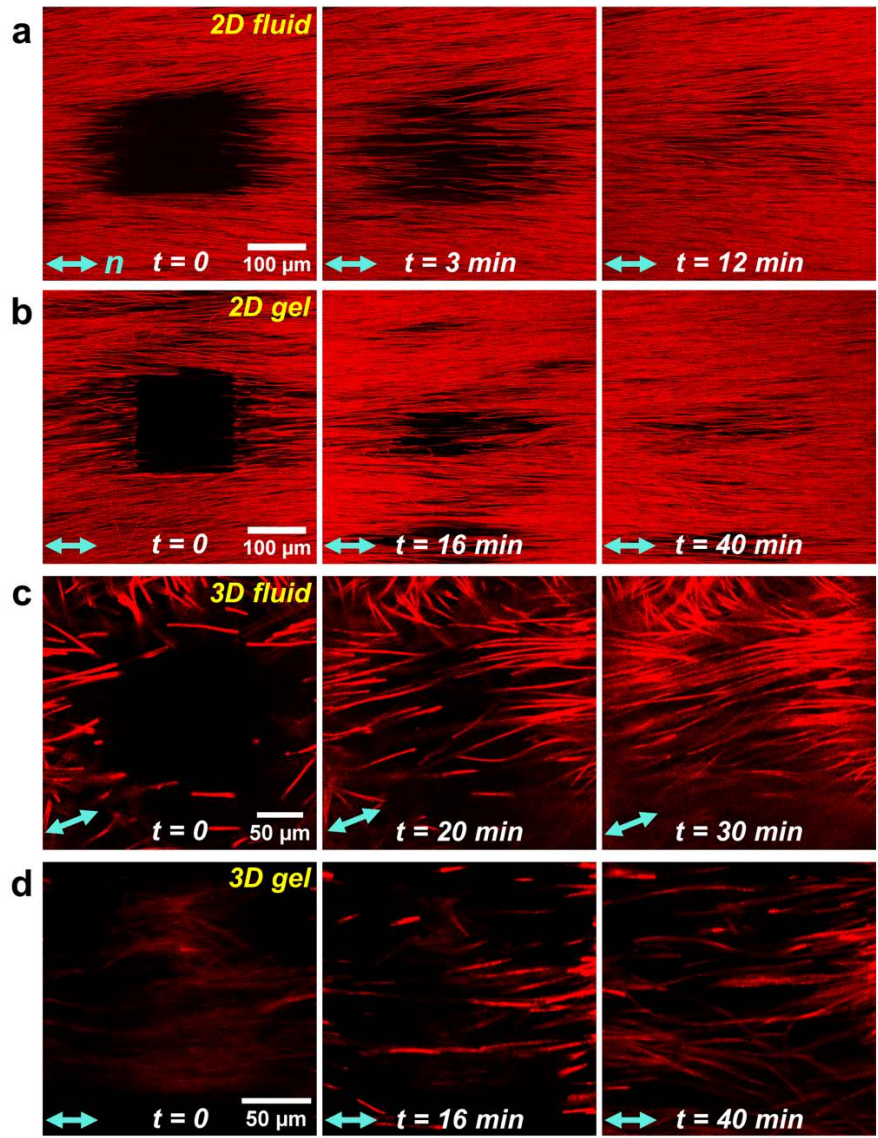

**Supplementary Fig. 5 | Recovery after photobleaching in active nematics assembled from *Geitlerinema* sp.**
**a-b,** Time evolution of a square-shaped photobleached monodomain region within (**a** and **b**) 2D nematic fluid and gel respectively, and (**c** and **d**) for 3D fluid and gel, respectively. Both 2D and 3D gels demonstrate sharper edges of the bleached regions as well as slower recovery to the uniform-intensity state as compared to their fluid counterparts.

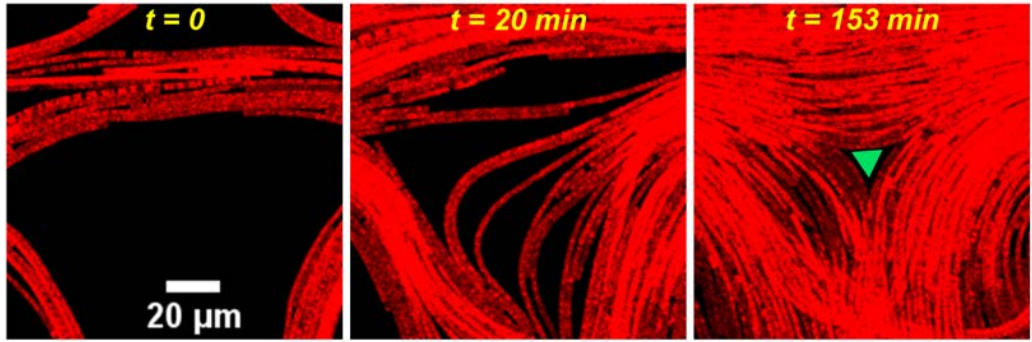

**Supplementary Fig. 6 | Optical images showing the emergence of a topological defect as nematic domains merge.**
The emergence of -1/2 defect in *Geitlerinema* sp. colonies observed when domains with different orientation meet, where the green triangle shown on the right-side frame indicates the defect core. Elapsed time is marke on the images.

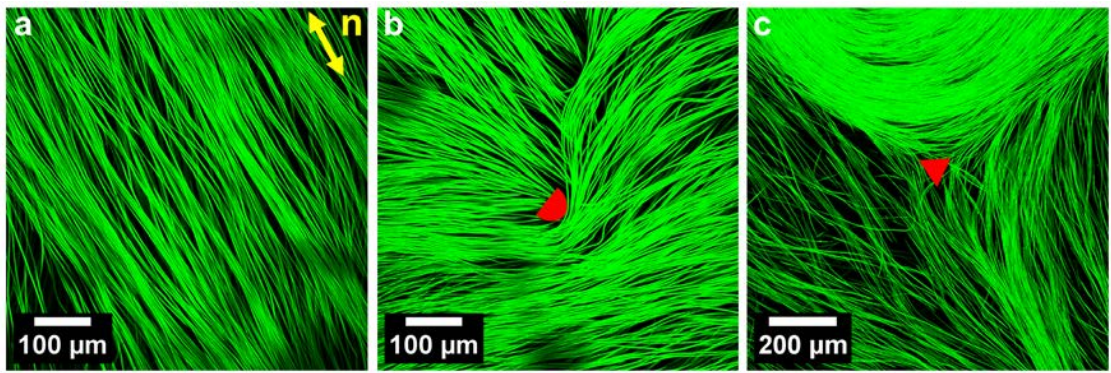

**Supplementary Fig. 7. Phototactic active nematic based on *Oscillatoria brevis*.**
**a-c,** Under controlled local illumination, *O. brevis* self-assemble into (**a**) a defect-free phototactic active nematic, where double arrow depicts the director **n**. Emergence of (**b**) +1/2 and (**c**) -1/2 defects, where red filled semicircle and triangle indicate the corresponding defect cores.

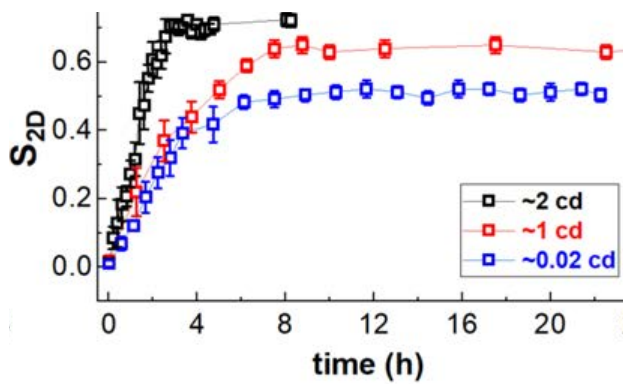

**Supplementary Fig. 8 | Intensity-dependent evolution of the orientational order parameter.**
Data show that phototaxis-induced transition from a disordered to an ordered state of phototactic cyanobacterial active matter under otherwise the same conditions takes longer times for low-intensity illuminations, with the corresponding long-time saturation of $S_{2D}$ at lower values.

# Captions for Supplementary Movies 1 to 10

**Movie 1. Collisions and coalignment of cyanobacterial trichomes.**

Interaction of motile straight *Geitlerinema* sp. filaments in a 2D sample which results in particle rotation and co-alignment. Yellow and cyan arrows are the eye guides for the self-propelled interacting rodlike trichomes behaving as polar particles.

**Movie 2. Diversity of filament trajectories and isotropic-nematic transition in 2D.**

Depending mainly on the *Geitlerinema* sp. trichome's length $L$, these bacteria reveal (**A**) circular trajectories, (**B**) straight trajectories, (**C**) self-propulsion with buckling. *O. brevis* filaments mainly reveal (**D**) straight traces; however, high-magnification video-microscopy € evidences the trichome rotation around its longer axis by tracking the filament's terminal cell. Typical videos illustrating the filament collective motility in (**F**) isotropic, (**G**) phototactic nematic fluid, and (**H**) phototactic nematic gel states in the locally illuminated area. In the concentrated nematic states (G-H), trichome propulsion becomes bipolar and parallel to **n**.

**Movie 3. 2D or quasi-2D motility in nematic fluids and gels imaged at the single cell level.**

High-magnification video-microscopy showing the 2D cyanobacterial motility in concentrated states with orientational ordering described by **n**: *Geitlerinema* sp. in nematic fluid (left), *Geitlerinema* sp. in nematic gel (middle), *O. brevis* in a nematic fluid state (right).

**Movie 4. Formation of a 2D nematic state at the water-air interface.**

Typical *Geitlerinema* sp. filament dynamics in the 2D nematic state emerging at the water-air interface. A few misaligned and out-of-focus filaments, which are visible in the video, swim in the water bulk underneath the 2D nematic domain.

**Movie 5. 3D isotropic-nematic and fluid-gel transitions.**

Continuous 3D isotropic-nematic transition taking place over the course of several hours, as observed for a *Geitlerinema* sp. cyanobacterial community. Volume reconstruction and typical X-Z cross sectional planes are shown on the right-side panel for isotropic (0-1.7 hours), nematic fluid (1.7-5.5 hours), and nematic gel (5.5-7 hours) states.

**Movie 6. Recovery after photobleaching for 2D and 3D nematic fluids and gels.**

Time series of images demonstrating *Geitlerinema* sp. filament recovery dynamics in 2D nematic fluid (top, left) and 2D nematic gel (top, right), as well as 3D nematic fluid (bottom, left) and 3D nematic gel (bottom, right) states. The microscopy observations reveal cyanobacterial dynamics after having the square-shaped regions photobleached with the laser beam.

**Movie 7. Defect dynamics in the 2D nematic fluid state.**

Time evolution of the director field near topological defects in a 2D phototactic nematic fluid state assembled from *Geitlerinema* sp.: the +1/2 defect core exhibits forward propulsion (left), whereas the −1/2 defect exhibits insignificant displacement (middle); the +1/2 defect core motion toward the −1/2 defect results in their annihilation. Corresponding defect trajectories are traced and shown by the green lines.

**Movie 8. Immobile defects in the 2D nematic gel state.**

Typical videos showing the static 2D nematic director field with the immobile defects in *Geitlerinema* sp. gel states. Filament motions around the +1/2 defect (left), and the −1/2 defect (middle), as well as +1/2 and −1/2 defect pair (right).

**Movie 9. 3D polydomain nematic gel.**

Large-scale video of *Geitlerinema* sp. based 3D phototactic nematic polydomain gel sample featuring filament motility along static tracks, as well as the immobile +1/2 and −1/2 disclinations.

**Movie 10. Foreign inclusions in a nematic gel.**

*Geitlerinema* sp. motility and emergence of quasi-boojum defects in a nematic gel caused by various foreign inclusions formed by crystalized salt.